\documentclass{aa}

\usepackage{graphicx}
\usepackage{subfigure}
\usepackage{txfonts}
\usepackage{epsfig}
\usepackage{natbib}
\bibpunct{(}{)}{;}{a}{}{,} 

\newcommand{\etal}{{\it et~al.~}}
\newcommand{\sax}{{\it BeppoSAX}}
\newcommand{\swf}{{\it Swift}}
\newcommand{\swx}{{\it Swift--XRT}}
\newcommand{\swu}{{\it Swift--UVOT}}
\newcommand{\swb}{{\it Swift--BAT}}
\newcommand{\xrt}{{\it XRT}}
\newcommand{\uvt}{{\it UVOT}}
\newcommand{\bat}{{\it BAT}}

\newcommand{\lat}{Fermi Gamma-ray Space Telescope-{\it LAT}}
\newcommand{\latf}{Fermi-{\it LAT}}
\newcommand{\tactic}{{\it TACTIC}}
\newcommand{\magic}{{\it MAGIC}}
\newcommand{\whipple}{{\it Whipple}}
\newcommand{\integi}{{\it INTEGRAL-ISGRI}}

\newcommand{\xmm}{{\it XMM-Newton}}

\newcommand{\asca}{{\it ASCA}}

\newcommand{\mrk}{Mrk~421}
\newcommand{\mrka}{Mrk~501}

\def\nh{$N_{\rm{H}}$~}

\begin{document}

\title{\swf~ observations of the very intense flaring activity of \mrk~ during 2006: I.
 Phenomenological picture of electron acceleration and predictions for the
MeV/GeV emission.}

\author{A.~Tramacere\inst{1,2}
	\and P. Giommi \inst{3}
	\and M. Perri \inst{3}
	\and F. Verrecchia \inst{3}
	\and G. Tosti \inst{4,5}
}

\institute{CIFS - Torino, Viale Settimio Severo 3, I-10133, Torino, Italy 
\and SLAC, 2575 Sand Hill Road, Menlo 	Park, CA 94025 USA,
\and ASI Science Data Center, c/o ESRIN, via G. Galilei, I-00044 Frascati, Italy\
\and Dipartimento di Fisica, Via A. Pascoli, I-06100 Perugia, Italy 
\and INFN Perugia, Via A. Pascoli, I-06100 Perugia, Italy
}

\offprints{tramacer@slac.stanford.edu}
\date{Received ....; accepted ....}

\markboth{A. Tramacere et al.:}
{A. Tramacere et al.:}

\abstract{}
{ 
We present results from a deep spectral analysis of all the \swf~ observations of  \mrk~ from April
2006 to July 2006, when it reached its largest X-ray flux recorded until 2006. The peak flux was about 85 milli-Crab
in the 2.0-10.0 keV band, with the peak energy ($E_p$) of the spectral energy
distribution (SED) laying often at energies larger than 10 keV. 
We study the trends among spectral parameters, and their physical insights in order to
understand the underlying acceleration and emission mechanisms.
}
{
We performed  spectral analysis of the \swf~ observations investigating the trends of the spectral parameters
in terms of acceleration and energetic features phenomenologically linked to the
SSC model parameters, predicting their effects in the $\gamma$-ray  band, in particular the
spectral shape expected in the \lat~ band.
}
{
We confirm that the X-ray  spectrum  is well described by a log-parabolic distribution close to
$E_p$, with the peak flux of the SED ($S_p$)  being correlated with $E_p$, and $E_p$ 
anti-correlated with the curvature parameter ($b$). 
The spectral evolution  in the  Hardness-ratio-flux  plane shows  both  clock-wise and counter-clock-wise patterns.
During the most energetic flares the UV-to-soft-X-ray spectral shape requires an electron distribution
spectral index $s\simeq 2.3$.
}
{
Present analysis shows that the UV-to-X-ray emission from \mrk~ is likely to be
originated by a population of  electrons that is actually curved, with a low energy power-law tail. The
observed spectral curvature is consistent both with stochastic acceleration or energy
dependent acceleration probability mechanisms, whereas the power-law slope form \xrt-\uvt~
data is   very close to that inferred from the GRBs X-ray afterglow and in agreement with the
\textit{universal}  first-order relativistic shock acceleration models. This scenario hints that the
magnetic turbulence may play a twofold role: spatial diffusion relevant to the first order process
and momentum diffusion relevant to the second order   process.
}

\authorrunning{A. Tramacere,\etal}

\titlerunning{  \swf~ observations of the very intense flaring activity of \mrk~ during 2006. }

\maketitle

\begin{table*}[htpb]
\caption{\swf~ observation journal and exposures of \mrk.}
\label{tab-log}
\begin{flushleft}
{\small
\begin{tabular}{llllll}

\hline

\noalign{\smallskip}

 ObsId      &Date     &Start UT   &\xrt~ Exp &\uvt~ Exp &\bat~ Exp\\
            &mm-dd-yy   &(s)        &(s)        &(s)     &(s)      \\ 
\hline
00206476000(*)  &04/22/06 (*) &04:21 AM  &10329   &10337  &15156\\
00030352005     &04/25/06  &06:22 AM  &4885    &1214   &4927\\
00030352006     &04/26/06  &03:29 AM  &3526    &878    &3567\\
00030352007     &04/26/06  &10:48 PM  &1328    &329   &1343 \\
00030352008     &06/14/06  &12:21 AM  &3187    &788   &3318 \\
00030352009     &06/15/06  &11:42 AM  &5427    &1336  &2787 \\
00030352010     &06/16/06  &12:33 AM  &23327    &5868   &23693 \\
00030352011     &06/18/06  &12:52 AM  &33288   &32468     &33671 \\
00030352012     &06/20/06  &11:59 PM  &15009   &0         &15379 \\
00030352013     &06/22/06  &01:08 AM  &20213    &20430     &0     \\
00030352014     &06/23/06  &09:25 AM  &7916    &0         &7221  \\
00215769000(*)  &06/23/06(*)   &03:44 PM  &1109    &1049   &5207 \\
00030352015     &06/24/06  &01:37 AM  &12944      &0         &13191 \\
00030352016     &06/27/06  &03:17 AM  &3046    &0         &3080 \\
00219237000(*)  &07/15/06(*)  &04:54 AM  &1916    &1915   &6224 \\
\hline
\noalign{\smallskip} 
\end{tabular}
}
\end{flushleft}
(*) Pointings triggered by the BAT instrument as GRBs.
\end{table*}

\begin{figure*}[]
\begin{center}
\epsfig{file=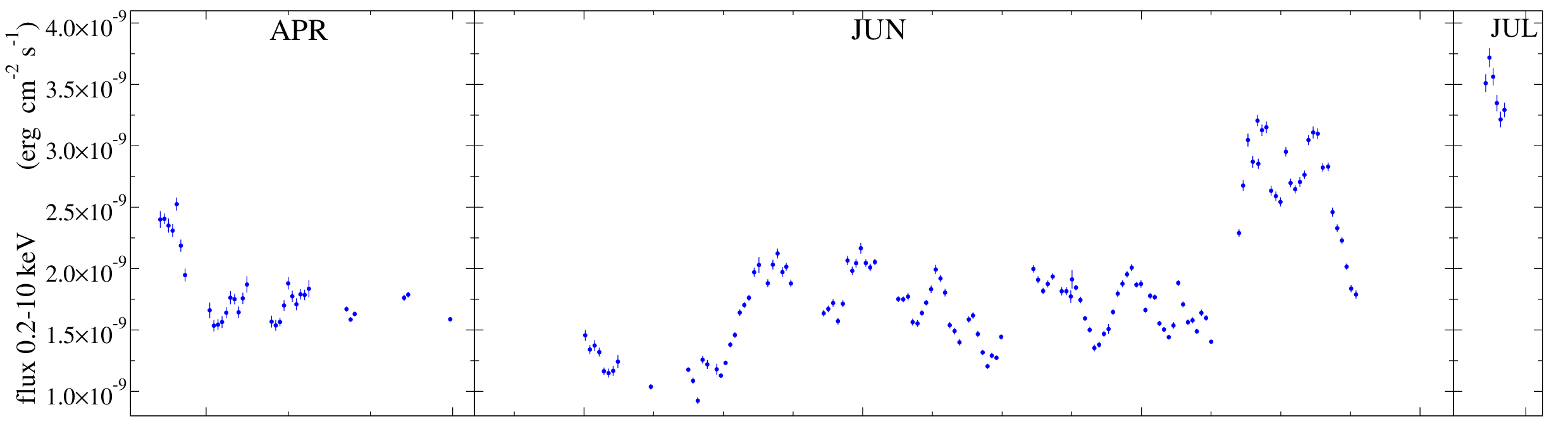,width=3.8cm,angle=-90}\\	
\epsfig{file=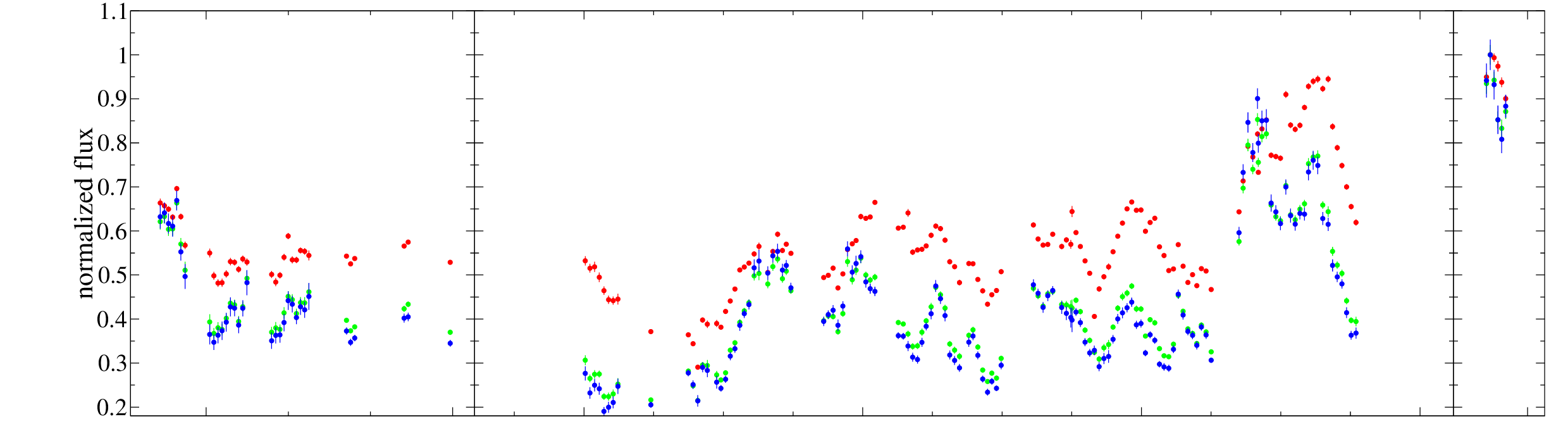,width=3.8cm,angle=-90}\\
\epsfig{file=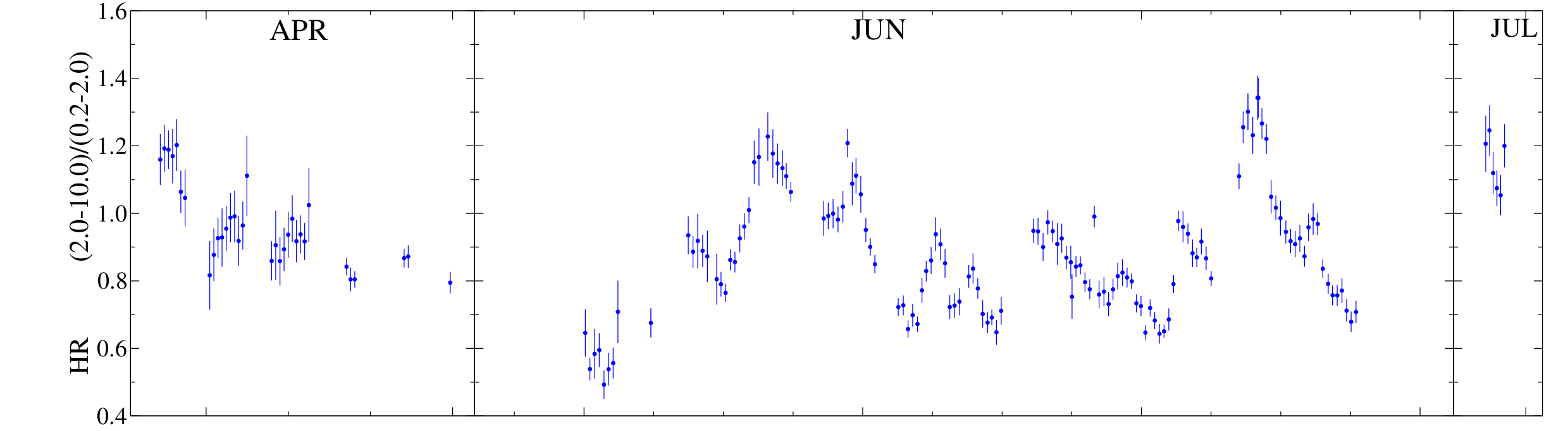,width=3.8cm,angle=-90}\\
\epsfig{file=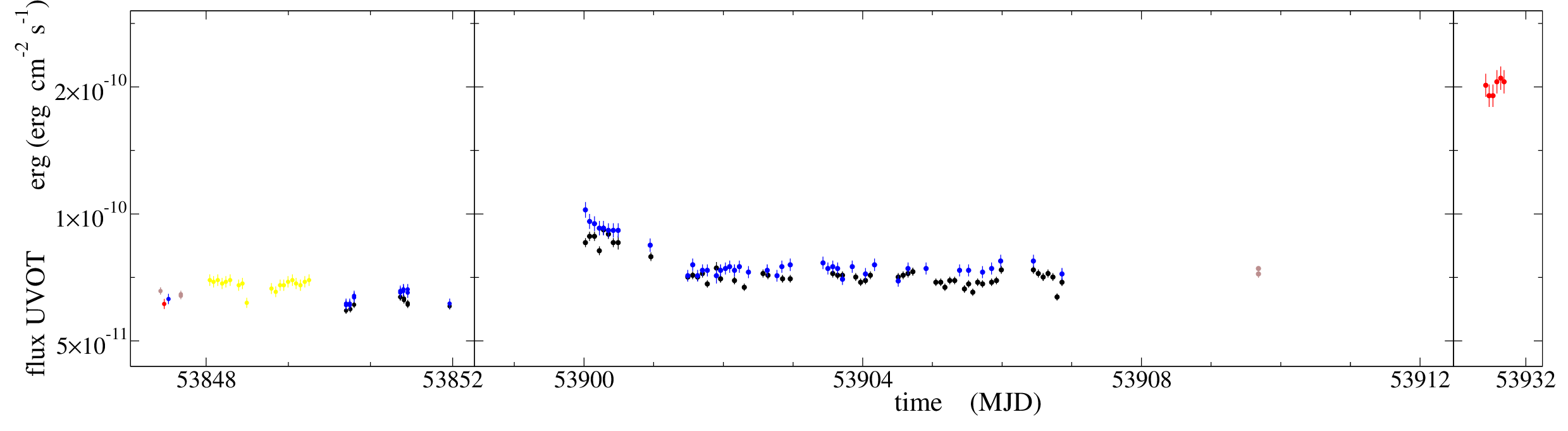,width=3.8cm,angle=-90}
\caption{ From top to bottom: a) light curve of the flux in the band $0.2-10.0$ keV. b) light curves
for three different bands soft ($0.2-3.0$ keV, red points) medium ($3.0-5.0$ keV, green points) and
hard ($5.0-10.0$ keV, blue points),
 normalized to their maximum value . c) the evolution of the hardness ratio (HR) evaluated as the
ratio of the $2.0-10.0$ keV band to the $0.2-2.0$ keV band. d) light curve from the \swu~ instrument,
different colors refers to different filters (V=brown, U=black, UVM2=red, UVW1=blue, UVW2=yellow).
[\textit{See the electronic edition of the Journal for a color version of this figure}]}
\label{multi-lc}
\end{center}
\end{figure*}

\section{Introduction}
BL  Lac objects are Active Galactic  Nuclei (AGNs)  characterized by a polarised 
and highly variable nonthermal continuum emission extending from radio to  $\gamma$-rays.
In the most accepted scenario this radiation is produced within a relativistic 
jet originated by the central engine and pointing close to our line  of sight. 
The   relativistic outflow has a typical bulk Lorentz factor $\Gamma \approx 10$, 
hence the emitted fluxes, observed at an angle $\theta$,  are subjected to the effects 
of  a beaming factor $\delta = 1/(\Gamma  (1 - \beta \cos\theta ))$. 

The Spectral  Energy Distribution  (SED) of these objects  has a typical 
two-bump  shape.  According to current models, the  lower-frequency bump 
is interpreted as synchrotron  emission from highly relativistic  electrons 
with Lorentz factors $\gamma$ in excess of $10^2$. This component peaks at frequencies 
ranging from the IR to the X-ray band.  The actual position of this peak has been 
suggested by \citet{Padovani1995}  as a marker for a classification;   
they define LBL (Low energy peaked BL Lac) the objects with the first bump in 
the IR--to--optical band, and HBL (High energy peaked BL Lac) those peaking in 
the UV--X-ray band. According to the Synchrotron Self Compton (SSC) emission mechanism,   
the  higher-frequency  bump is to be attributed to inverse Compton scattering 
of synchrotron photons by the same population of relativistic electrons that 
produce the  synchrotron emission \citep{Jones1974, Ghisellini1989}.

With its redshift $z$ = 0.031, \mrk~  is among the closest and best studied HBL. 
In fact, it is one of brightest BL Lac objects in the UV and in the X-ray bands, 
observed in $\gamma$ rays by EGRET \citep{Lin1992}; it was also the first extragalactic 
source detected at TeV energies in the range 0.5-1.5 TeV by the Whipple 
telescopes \citep{Punch1992, Petry1996}.

The source is classified as HBL because its synchrotron emission peak ranges  
from a fraction of a keV to several keV. Its flux changes go along with strong 
spectral variations \citep{Fossati2000a,Massaro2004a} and the spectral shape 
generally exhibits a marked  curvature, well described  by  a log-parabolic model 
\citep{Massaro2004a, Trama2007b}.

In spring/summer 2006  \mrk~  reached its largest X-ray flux recorded until 
that time. The peak flux was about 85 milli-Crab in the 2.0-10.0 keV band, with 
the peak energy of the spectral energy distribution (SED) often  lying at energies 
larger than 10 keV.\\ 
In this paper (Paper I) we present  data collected  from \swf~ observations 
performed during this very intense flaring period. 
 We aim to study the evolution of the spectral parameters as a function of the 
flaring  activity, and the correlations among the spectral parameters. This gives  
a phenomenological picture of the physical mechanism driving the observed patterns.  
In Paper II we will frame this scenario in the theoretical context of stochastic
acceleration  \citep{Tramacere2009}.\\

In the phenomenological context of jets in HBLs the spectral curvature 
is  relevant for understanding of both radiative and acceleration
mechanisms.\\ 
Many works in the literature have shown that the X-ray spectral shape of \mrk~ 
is actually curved and described by a log-parabolic distribution with a mildly 
curved and symmetric spectral shape 
\citep{Fossati2000b,Tanihata2004,Massaro2004a,Trama2007a}. 
\citet{Massaro2004a} gave an interpretation of this feature in the framework of
energy dependent acceleration efficiency that naturally leads  to log-parabolic 
spectral distributions with a possible power-law tail at lower energies. 
\cite{Karda1962} showed that a log-parabolic distribution results from a 
stochastic acceleration scenario with a mono-energetic or quasi-mono 
 energetic particle injection.
\citet{Katar2006} and \citet{Giebels2007} used a relativistic Maxwellian electron 
distribution, resulting from a stochastic acceleration process, to describe the 
X-ray/TeV emission of \mrka~ and \mrk~ respectively. 
Recently, \citet{Staw2008} showed that a distribution with  similar spectral 
properties can be obtained as a steady state energy spectra of particles undergoing 
momentum diffusion due to resonant interactions with turbulent MHD modes.\\

\citet{Trama2007b} suggested that the connection of the X-ray  curvature with 
 that in the emitting particles  and its evolution with the source state, 
 could be investigated as a test for the prediction of the scenarios listed above.  
In particular, both stochastic \citep{Karda1962} and energy-dependent 
acceleration mechanisms predict  an anticorrelation between  the curvature 
and the SED peak energy ($E_p$).\\
Moreover, the pattern shown by the peak height ($S_p$) of the SED as $E_p$ moves, 
can trace the evolution of the parameters characterizing the energetics 
of the synchrotron emission, in particular the average particle energy  and the 
number density of the emitting particles.\\

A crucial issue is understanding whether during the most violent flares the 
shape of the X-ray spectrum also can be described by a single log-parabolic spectral 
distribution. In fact,  a typical X-ray detector  shows only a slice 
(usually up to  two decades) of the overall emission from the observed object. 
During strong flares involving large variation of the SED peak energy, it is possible 
to understand if the electron distribution is curved even far from the peak energy. 
Moreover, we can capitalise on the unique opportunity given by \swf~ to perform 
simultaneous UV-to-X-ray observations, extending the spectral window from about 
$10^{15}$ Hz up to $10^{19}$ Hz. The presence of a power-law tail at low photon 
energies and its slope can provide information about the low-energy tail of the 
underlying  electron distribution as well as on the acceleration mechanism 
generating such a spectral shape.\\

As final task,  building on the phenomenological results from the present analysis, 
we can model the SED of \mrk~ within the synchrotron-self-Compton (SSC) 
scenario, predicting the possible spectral behaviour at 
$\gamma-$ray energies. In particular we can relate the typical spectral shape
 of the UV-to-soft-X-ray emission to that expected in the MeV/GeV band covered by the
\lat~ instrument.\\

This paper is organized as follows. In Sect. 2 we present our data set and the 
procedure used to reduce the \swf~ data. In Sect. 3 we report results from the
analysis of the fast variability of the source. In section 4 we present spectral analysis 
results. In section 5 we analyse the spectral evolution of the source. In Sect. 
6 we study the trends among the spectral parameters and compare these results with
previous studies  directing our attention to the link between such trends and
expectations from different  scenarios for acceleration mechanism.
In Sect. 7 we study the connection between the \uvt~ and the \xrt~ spectra, 
showing the relevance of the derived spectral shape in the context of the first 
order acceleration processes. In sect. 8 we model the SED within the SSC framework, 
focusing on the phenomenological interpretation of the data. In Sect. 9 we 
discuss the  results, and in Sect. 10 we  draw our overall conclusions.

\section{Swift Observations and data analysis}
We present results from temporally-resolved spectral analyses of 15 \swf~  
simultaneous  observations in the UV/X-ray band performed between April and July 
2006, when the source was so bright that the high-energy instrument $BAT$ 
automatically triggered pointings at it  three times assuming as if it were a 
Gamma Ray Burst.
The log of UV/X-ray observations is reported in Tab. \ref{tab-log}.\\
In  Fig. \ref{multi-lc} we report the \xrt~   light curves  obtained from the 
single-orbit spectra.  In the first panel from the top  we show the light curve 
of the flux obtained  integrating the model from Eq. \ref{eq-lp} (see section 4.2)
between 0.2 and  10.0 keV, according to the parameters  and parameters errors 
reported in Tab  \ref{tab-fit}.
Fluxes in Tab. 2  refer to the 0.3-10.0 keV interval because  the \xrt~ response 
function is calibrated only in that range. We extrapolated  the flux to 
the 0.2-10.0 keV  band  to make easier  comparisons with data from other X-ray 
telescopes  that usually are given in the 0.2-10.0 keV band.\\ 
The second panel from the top shows the light curve for three different bands: 
soft ($0.2-3.0$ keV) medium ($3.0-5.0$ keV) and hard ($5.0-10.0$ keV), normalized 
to their maximum value.  In the third panel from the top, we report the evolution 
of the hardness ratio (HR) evaluated as the ratio of the $2.0-10.0$ keV band to 
the $0.2-2.0$ keV band.\\
The bottom panel of Fig. \ref{multi-lc} shows the light curve obtained from 
the \swu~ observations.\\



\subsection{\swx~ data analysis }
All the data were reduced using the $XRTDAS$ software (version v2.2.0) developed 
at the ASI Science Data Center (ASDC) and distributed within the HEAsoft 6.4 package 
by the NASA High Energy Astrophysics Archive Research Center (HEASARC).

The operational mode of \xrt~ is automatically controlled by the on-board software 
that uses the appropriate CCD readout mode to reduce or eliminate the effects of 
photon  pile-up.  
When the target count-rate is higher than $\approx$1 cts/s the system is normally 
operated in Windowed Timing (WT) mode whereas, the Photon Counting (PC) mode is 
used for fainter sources \citep[see][for more details on \xrt~ observing modes]{Burrows2005,Hill2004}. 
The observations presented here were all performed in WT mode, owing to the 
extremely high state of the source  (40-80 cts/s) .\\
We selected photons with grades in the range  0--2 for WT mode; we also used
default screening parameters to produce  level 2 cleaned event files. To take 
advantage of the good statistics offered by such large numbers of events 
we decided to make a  high temporal resolved  analysis, extracting spectra for 
each \swf~ orbit. We  rejected only two out of the 174 obtained spectra, because 
of a strongly biased exposure.  The resulting \xrt~ database  presented in this 
work therefore includes 172 time intervals.

\subsection{\swu~ data analysis }
The Swift UV and Optical Telescope  \citep[UVOT,][]{Roming2005} observations included
 in this paper have exposures in all filters except for the White one.
Photometry of the source was performed using the standard UVOT software developed 
and distributed within the HEAsoft  6.4 package. Counts were extracted from an 
aperture of 5\arcsec\, radius for all single exposures within an observation and 
for all filters, while the background was carefully estimated in few ways.
In almost all observations the source is on the ``ghost wings'' \citep{Li2006} 
from the nearby star 51 UMa, so we estimated the background from a circular aperture 
of 15\arcsec\ radius off the source but on the wings, excluding stray light and 
support shadows. These background values were compared to those obtained using a 
region outside the wings, resulting in differences in most cases within the errors. 

We checked astrometry of each exposure, verifying the aperture positioning.
 Count rates were then converted to fluxes using the standard zero points. We 
discarded some exposure for which the count rate was near the limit of acceptability 
for the ``coincidence loss'' correction factor included in the CALDB 
($\sim$\,90 {\rm cts s}$^{-1}$). 

The fluxes were then de-reddened using a value for $E(B-V)$ of 0.014 mag 
\citep{Schlegel1998} with $A_{\lambda}/E(B-V)$ ratios calculated for UVOT filters 
(for the latest effective wavelengths) using the mean Galactic interstellar 
extinction curve from \cite{Fitz1999}.

\subsection{\swb~ data analysis }
We analysed the data that  automatically triggered the \bat~ instrument  (labeled with  *) 
in Tab. \ref{tab-log}). To reduce the data we followed the instructions reported 
in the \bat~ analysis threads \texttt{http://swift.gsfc.nasa.gov/docs/swift/analysis/
threads/batspectrumthread.html}. We first generated a detector plane image (DPI) 
and checked for \textit{noisy} pixels using the \texttt{bathotpix} task. In order 
to properly subtract the background we generated a weighting mask  using the 
\texttt{batmaskwtevt} task. The spectrum was extracted using the \texttt{batbinevt} 
task. We applied geometric corrections (\texttt{batupdatephakw}), and systematic 
error correction(\texttt{batphasyserr}). As final task we generated a response 
matrix (\texttt{batdrmgen}). The spectrum was rebinned by providing an external
bin edges file to \texttt{batbinevt}.

\section{Fast source variability.}
\begin{figure}[!h]
\epsfig{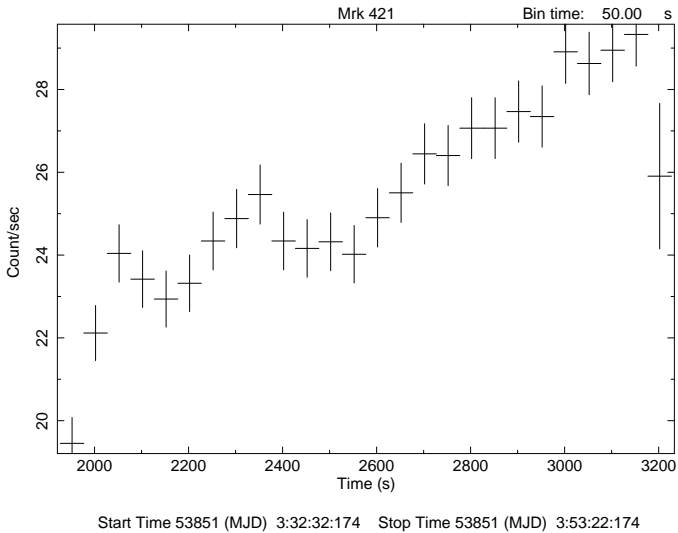}
\caption{ Light curve binned with 50 s intervals, 
 from the first orbit of the 26/04/2006 pointing. Significant variability ($\simeq 35\%$)  
is detected over intra-orbit time scale (1200 s).}\label{fig-fv}
\end{figure}

Temporal variability is one of the most interesting features characterizing HBLs.
In the X-ray band   typical time scales for flux changes decrease to the order 
of hour or minutes. The identification of the order of the fastest significant flux 
change  allows to estimate the source size, given by the well-known relation:
\begin{equation}
\label{size}
R \leq \frac{c~\Delta t~\delta}{1+z} ,
\end{equation}
where $c$ is the speed of the light, $\delta$ the beaming factor, and $z$ the 
cosmological redshift. We investigated the intra-orbit light curves from our data 
set and found a significant  flux variation  in the first orbit on 04/26/2006 
(ObsID 00030352006) pointing (see Fig. \ref{fig-fv}). The light curve extracted 
from the $0.3-10.0$ keV band shows an increasing trend spanning  about $ 1200$ 
seconds with a variation of  about $35\%$.\\

According to Eq. \ref{size}, a  $1200$ s time scale implies $R\leq 3.6\times
10^{13} \delta$ cm. Assuming a beaming factor of the order of 10 we get $R \simeq
4\times10^{14}$ cm, which indicates a quite compact emitting region.\\
\citet{Lich2008} analysed the temporal variability of \mrk~ using observations 
performed by  the \integi~ instrument  in the $40-100$ keV band, and overlapping 
our data set during the June pointings.  The fastest variability observed in 
the \integi~ light curves, estimated by fitting the data with  a rise-time law
of the form $a\cdot e^{t/t_0}$, gives $R \lesssim 3\times 10^{14} \delta $ cm.
We perform the same analysis for the  04/26/2006 light curve and we find $t_0\simeq 4.100$ s 
which implies $R \lesssim 1\times 10^{14}\delta$ cm.  We use this time scale to 
constrain the emitting region size in the following analyses (Sect. 8).\\
Time scales similar to that  in the 04/26/2006 pointing  were also observed by 
\citet{Tanihata2001} in  \asca~ data,  about $5$ ks 
($R \lesssim 1.5\times 10^{14}\delta$ cm), and by \citet{Giebels2007} 
(about 2  ks) analysing TeV data from the CAT telescope 
($R \lesssim 5.4\times 10^{13}\delta$ cm).\\ 

\section{Spectral analysis}

\subsection{\xrt~ spectral analysis}
\begin{figure*}[]
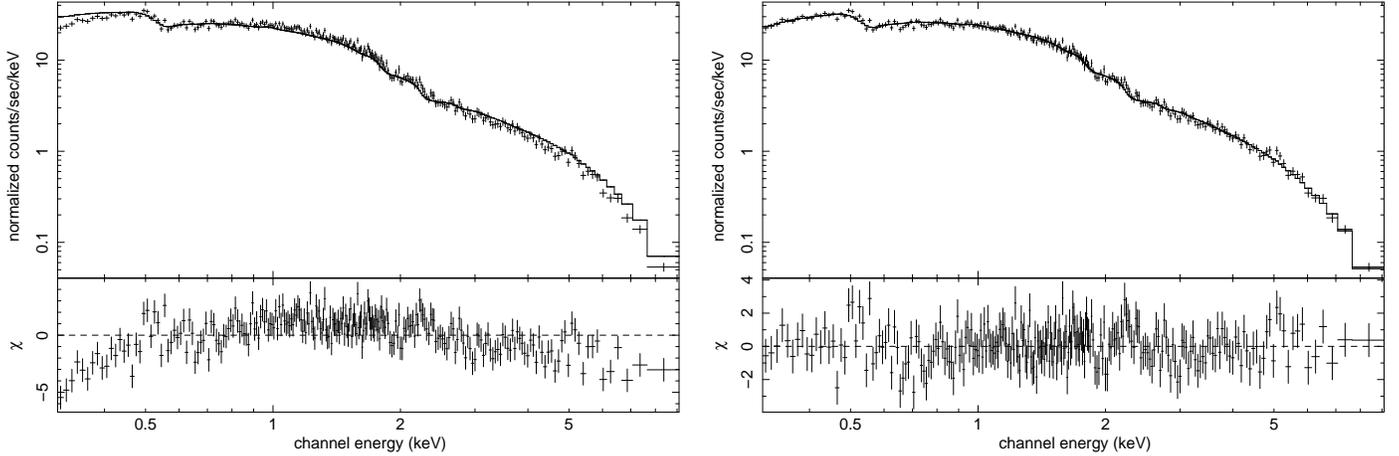

\begin{tabular}{ll}
\epsfig{file=pl.ps,width=6cm,angle=-90}
&\epsfig{file=lp.ps,width=6cm,angle=-90}\\
\end{tabular}
\caption{Spectrum from the first orbit  of the ObsID 00030352013 performed
on 06/22/2006. \textit{Left Panel}: the systematic deviations on both sides of
the residuals from a best fit with a power-law with Galactic \nh~ show the need of intrinsic
curvature. \textit{Right Panel}: the deviations disappear with the log-parabolic model with
Galactic \nh . The $\chi^2_{r}$ decreases from 1.60 with 246 d.o.f. (power-law) to 1.19 with 245  d.o.f.
 (log-parabola); the F-test statistics clearly favours the curved model.}\label{fig-pl-fit}
\end{figure*}

We find, for most of the spectra in our sample, 
systematic deviations (see Fig. \ref{fig-pl-fit}) in the residuals
obtained when	 fitting  the data by means of  a power-law spectrum with 
\nh~ fixed at  the Galactic value. Such a behaviour heuristically suggests 
that the spectra are intrinsically curved. This was already known in the literature
\citep{Fossati2000b,Tanihata2004,Massaro2004a,Trama2007a,Trama2007b}. 
All these authors agreed  that when the spectral shape of \mrk~	 is curved,  
describing the curvature only in terms of absorption not only would require a 
column density much higher than the Galactic value $N_H=1.61\times 10^{20}$cm$^{-2}$ 
\citep{Lockman1995}, but also would yield in any case unacceptable fits with very 
high $\chi^2_r$. Moreover,  high resolution images of the host early-type galaxy 
of \mrk~ do not show in the brightness profile any evidence of large amounts of 
absorbing material \citep{Urry2000}. Based on these phenomenological results,  
we performed the spectral analysis fixing the \nh~  absorbing column densities to 
the Galactic values and using the following  log-parabolic spectral
law (LP):\\
\begin{equation}
 F(E) = K~E^{-(a+b~\log(E))} ~~~~~~ \mbox{ ph cm$^{-2}$ s$^{-1}$ keV$^{-1}$},
\label{eq-lp}
\end{equation}
where $a$ is the photon index at 1 keV   and $b$ measures the spectral
curvature. \\

The SED peak energy  ($E_p$) and  the SED height ($S_p$) can be  derived 
easily from  Eq. \ref{eq-lp}, but in this case they suffer in intrinsic analytical correlation. 
This bias can be removed using an equivalent functional relationship that is a 
log-parabola expressed in terms of $E_p$, $S_p$ and $b$ (LPEP):
\begin{equation}
S(E)= (1.60 \times 10^{-9})~S_p~10^{-b~(\log(E/E_p))^2}~~~~~~\mbox{ erg cm$^{-2}$ s$^{-1}$},
\label{eq-lpep}
\end{equation}
where $S_p=E_p^2 F(E_p)$  and  $E_p$   are estimated  during the fit, and the 
numerical constant is simply the energy conversion factor from keV to erg.\\

\subsection{Orbit resolved analysis}
Because of  the very bright state of the source, we were able to extract spectra 
for each orbit, for a total of 172 spectra. A motivation to perform an orbit resolved 
analysis is the strong variability of the source during these pointings. 
In fact, integrating spectra over time scales much longer than the typical variability 
leads to misleading results in the estimates of the curvature, $E_p$, and $S_p$.\\

The results of the spectral analysis are reported in Tab. \ref{tab-fit} 
(rejected spectra are labeled with (*)), where all statistical
errors refer to the $68\%$ confidence level (one Gaussian standard deviation).
The second thorough  forth columns in Tab. \ref{tab-fit} report the best fit 
estimates for the model in Eq. \ref{eq-lp}. The fifth column reports the value of 
the SED peak analytically  estimated from Eq. \ref{eq-lp} according to the best 
fit results ($E_p*$ ). The sixth and seventh columns report the $E_p$ and $S_p$ best fit 
estimates using as  fitting  model Eq. \ref{eq-lpep}. In the eighth column
we report the flux in the 0.3-10.0 keV band, evaluated by X-spec integrating the 
Eq. \ref{eq-lp} model. In the last column we report the reduced $\chi^2$ statistics 
for the fit with  Eq. \ref{eq-lp}.\\

The SED peak energy was often difficult to  estimate. The reason was that 
during this particular high brightness state, the spectra were in some cases very 
hard, with a photon index $a \simeq [1.6-1.7]$ and and having a low spectral 
curvature,  implying a peak energy placed far out of the \xrt~ energy band.\\ 
In order to  test the robustness of the $E_p$ estimate, we first derived the peak 
energy from  the spectral parameters of Eq. \ref{eq-lp} ($E_p*$). Then we fitted 
the spectra using Eq. \ref{eq-lpep}, setting the initial curvature value to that 
returned from the fit with Eq. \ref{eq-lp}. In order to test the stability of the 
results  we adopted the following criteria:
\begin{enumerate}
 \item The value of $E_p$   statistically significant. Given the asymmetric
uncertainties we define  $\sigma_{E_p}$ the half 2 sigma confidence level, and 
require  $E_p/\sigma_{E_p}<1$.
\item $E_p*$ consistent with $E_p$ at within one sigma.
\end{enumerate}
We report in Tab. \ref{tab-fit} the estimates of $E_p$ satisfying this criterion, 
in the other cases we report only the lower limit of $E_p*$.
The estimates of $E_p*>100$ keV obviously are not statistically robust, meaning 
that the actual energy peak may be in excess of $100$ keV, but we could not give 
a robust estimate.\\
All the spectra for which  the stability conditions were satisfied returned 
values of $E_p \lesssim 20$ keV.\\ 

\subsection{Orbit merged analysis }
The orbit resolved spectral analysis has the great advantage to follow accurately 
the strong variability of the source, but  the $E_p$ estimates suffer from large 
uncertainties. In any case, based on the spectral/flux pattern traced by the previous 
analysis, we can identify all the orbits indicating essentially the same 
spectral/flux states.
We can use these intervals to perform an orbit-merged spectral analysis, 
to get smaller uncertainties on the $E_p$ value, and without integrating  the 
source over periods exhibiting large changes.\\
The result of this analysis is reported in Tab. \ref{tab-fit-1}. 
In this analysis when $E_p$ and $E_p*$ can be determined the typical 
uncertainties are smaller. We  use the values from  this table to perform the
 $E_p-S_p$ and $E_p-b$ trend analysis in the following, selecting only the data 
with significant $E_p$  estimates. Also in this case, all the spectra with $E_p$ 
well constrained have $E_p < 20$ keV.

\begin{figure}[]
\epsfig{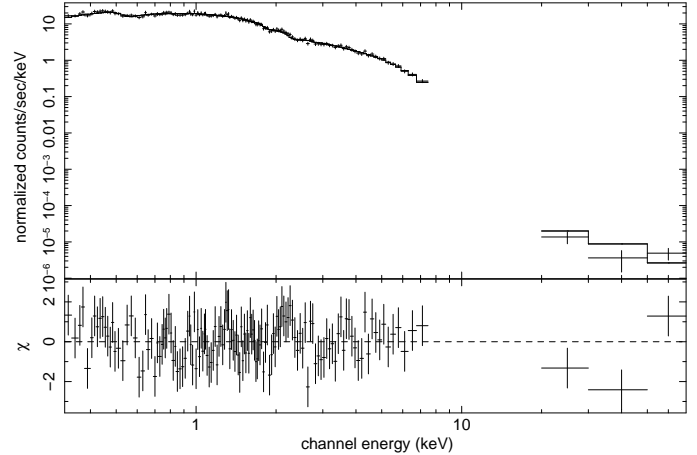}
\caption{Joint \xrt~ \bat~ spectral analysis from the 04/22/2006 pointing.}\label{fig-x+b-fit}
\end{figure}

\subsection{Joint \xrt-\bat~ spectral analysis}
For the three observations that have simultaneous \swx~ and \swb~ data
(with the \bat~ in automatic trigger mode, see Tab. \ref{tab-log}) we performed 
a joint \xrt-\bat~ spectral analysis. The results are reported in Table 
\ref{tab-fit-1} (lines labeled X+B).
The spectral curvature resulting from the joint analysis is slightly larger with 
respect to to the case of only \xrt~ data, but is always consistent 
within one sigma. As already discussed, the estimation of $E_p$ is more subtle. 
We will use the  superscript $^{X+B}$ to refer to joint \xrt~ and \bat~ 
analysis hereafter.\\
For the 07/15/2006 pointing, the two curvatures are $b=0.17\pm0.02$ and
$b^{X+B}=0.20\pm0.02$, and the two values of $E_p$  consistent within one sigma 
are $E_p=11_{-2}^{+4}$ keV  and $E_p^{X+B}=8^{+2}_{-1}$ keV.  
For the case of the 04/22/2006 pointing (Fig. \ref{fig-x+b-fit}), 
the peak energies resulting are $E_p^{X+B}=20^{+10}_{-6}$ keV and 
$E_p=26^{+19}_{-8}$ and the curvatures $b=0.11\pm0.02$ and $b^{X+B}=0.12\pm0.02$.\\
For the 06/23/2006 pointing, the \xrt~ curvature is estimated $0.08\pm0.03$ to 
be compared to the value of $b^{X+B}=0.13\pm0.02$. In this case the smaller value 
of the curvature, as discussed in the previous section, makes more difficult the 
estimate of the $E_p$ value. In fact, \xrt~ data are not able to locate the actual 
value of the peak energy, and the estimate from joint analysis is 
$E_p^{X+B}=34^{+22}_{-11}$ keV.  \\

The results  from joint \xrt-\bat~ analysis  confirm that $E_p$ values 
estimated  from \xrt~ data, at energies larger than about 20 keV, are potentially 
biased. The selection applied in the present analysis should warrantee that the  
bias on the trends among $E_p$ , $b$ and $S_p$ will be as small as possible.

\begin{figure}[]
\epsfig{file=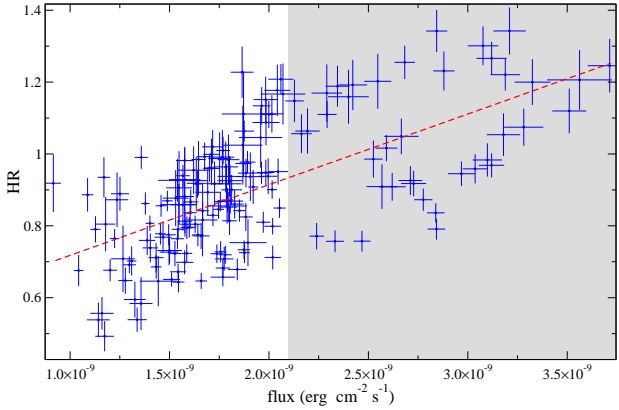,width=5.5cm,angle=-90}
\caption{Scatter plot of the HR vs. the flux in the 0.2-10 keV band (HR is evaluated as the ratio
of the $2.0-10.0$ keV band to the $0.2-2.0$ keV band). The points
in the shaded area are almost all from the period 06/23/2006 to 06/27/2006 and
07/15/2006.}\label{fig-HR-flux} 
\end{figure}

\section{Spectral evolution}

\begin{figure*}[]
\begin{center}
\begin{tabular}{ll}
 
\epsfig{file=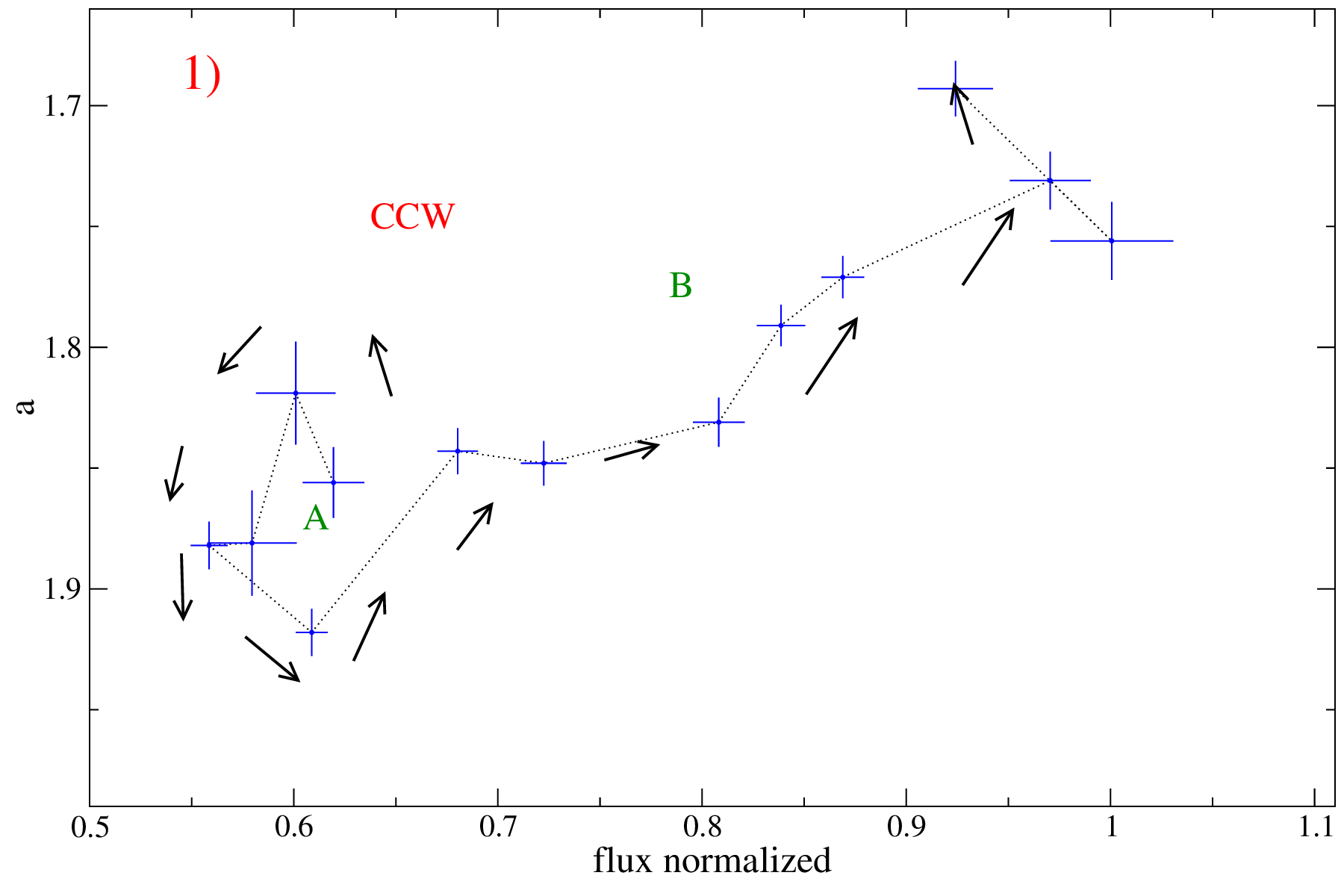,width=5.5cm,angle=-90}& 
\epsfig{file=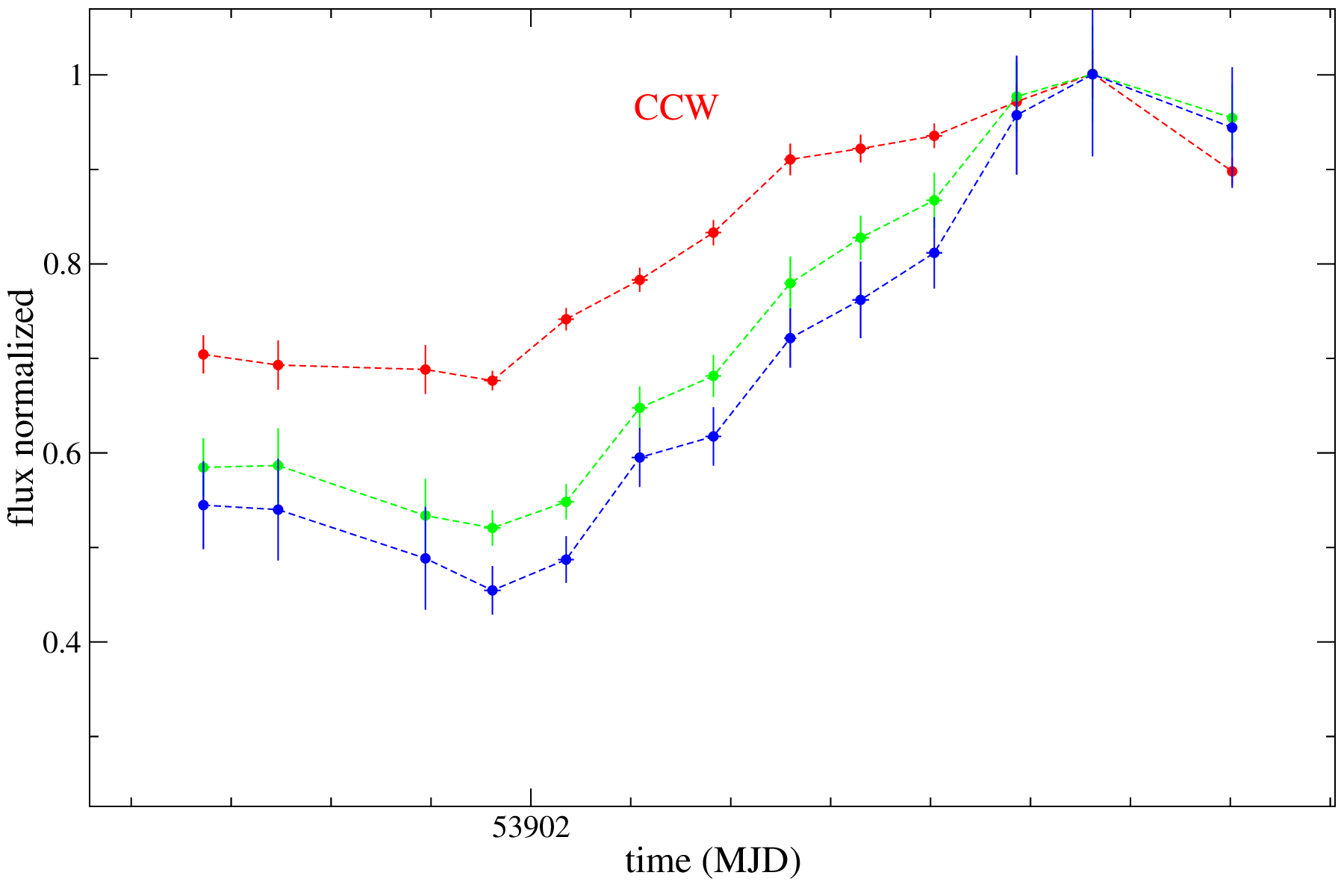,width=5.5cm,angle=-90}\\
 
\epsfig{file=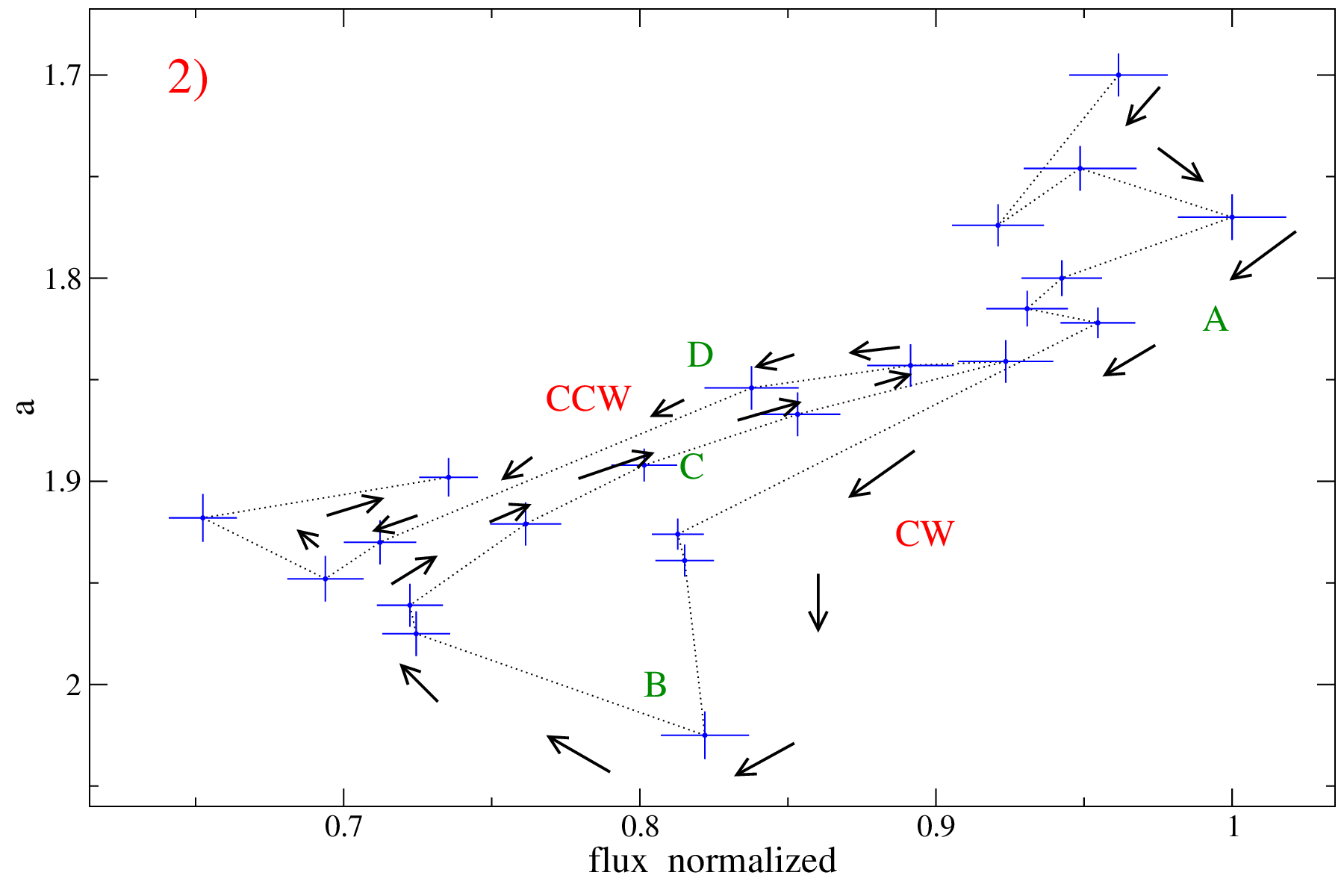,width=5.5cm,angle=-90}&
\epsfig{file=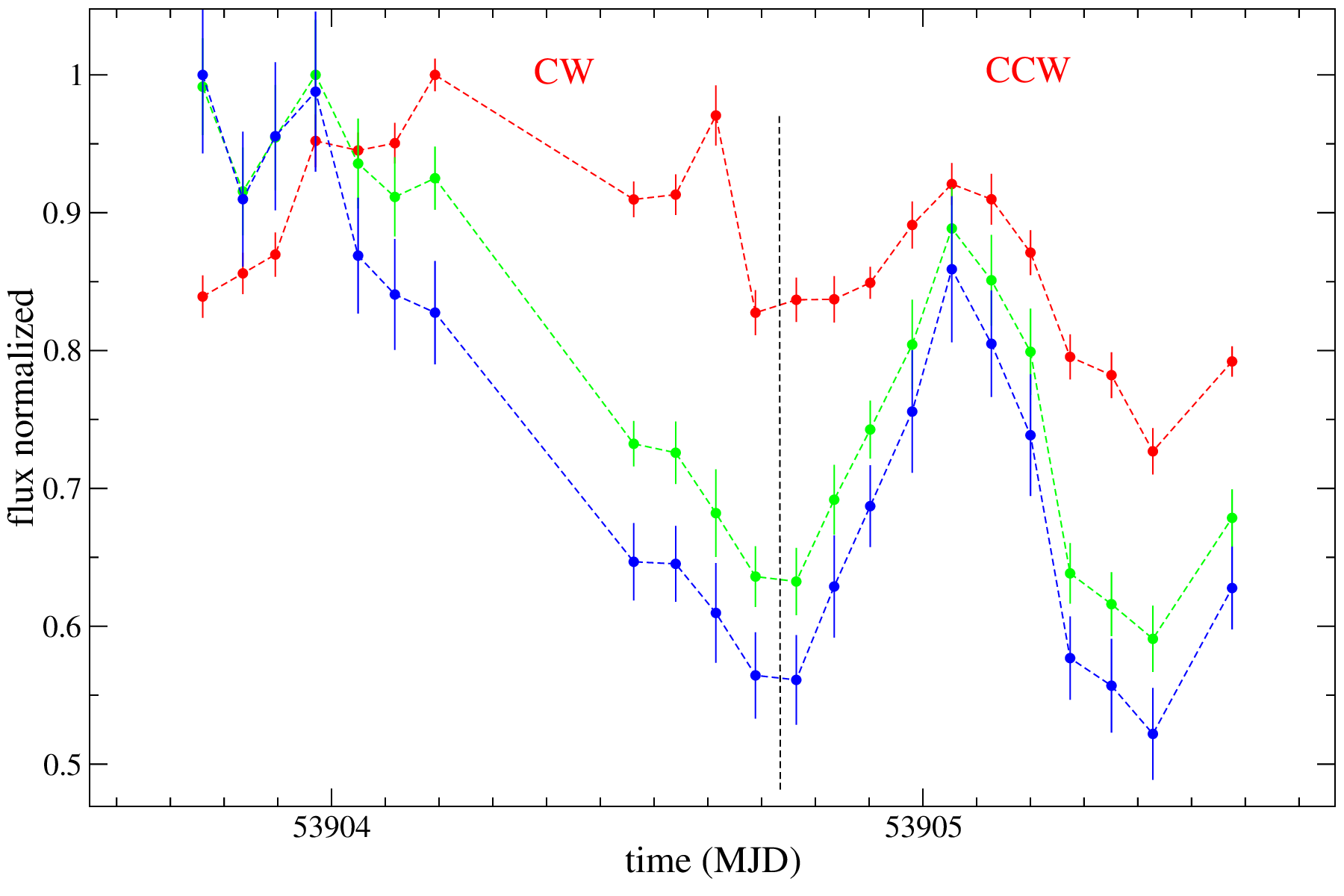,width=5.5cm,angle=-90}\\

\epsfig{file=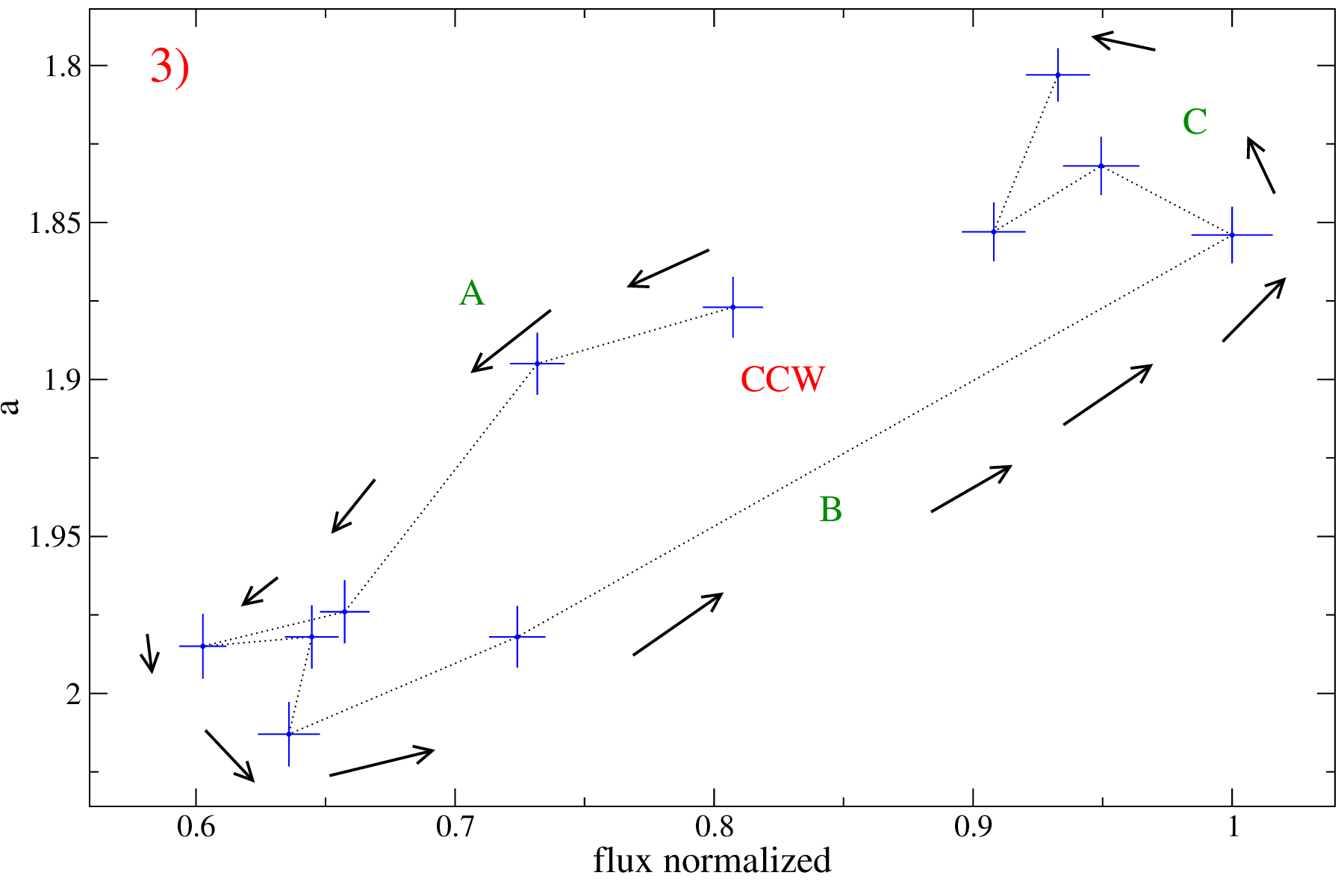,width=5.5cm,angle=-90}&
\epsfig{file=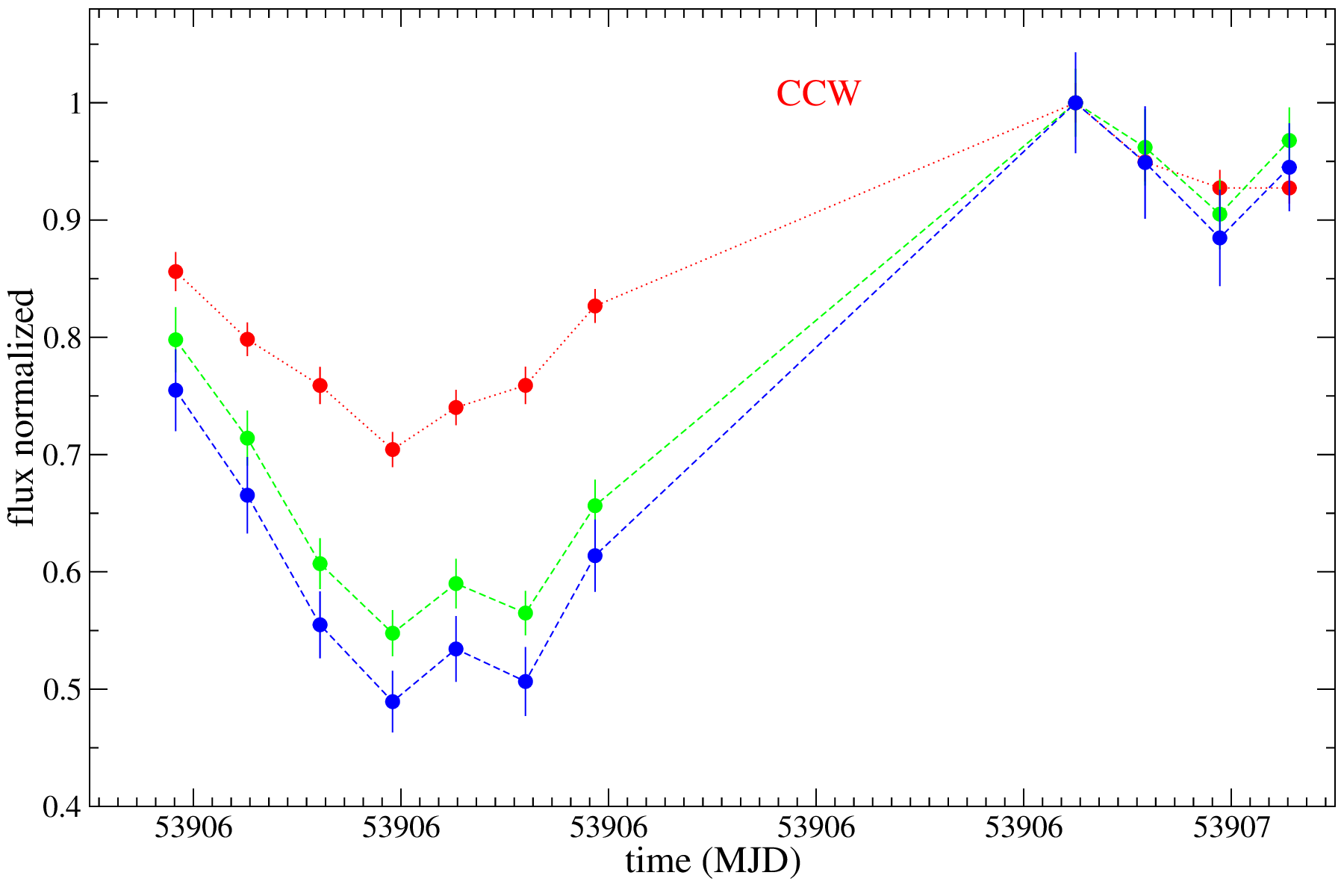,width=5.5cm,angle=-90} \\

\epsfig{file=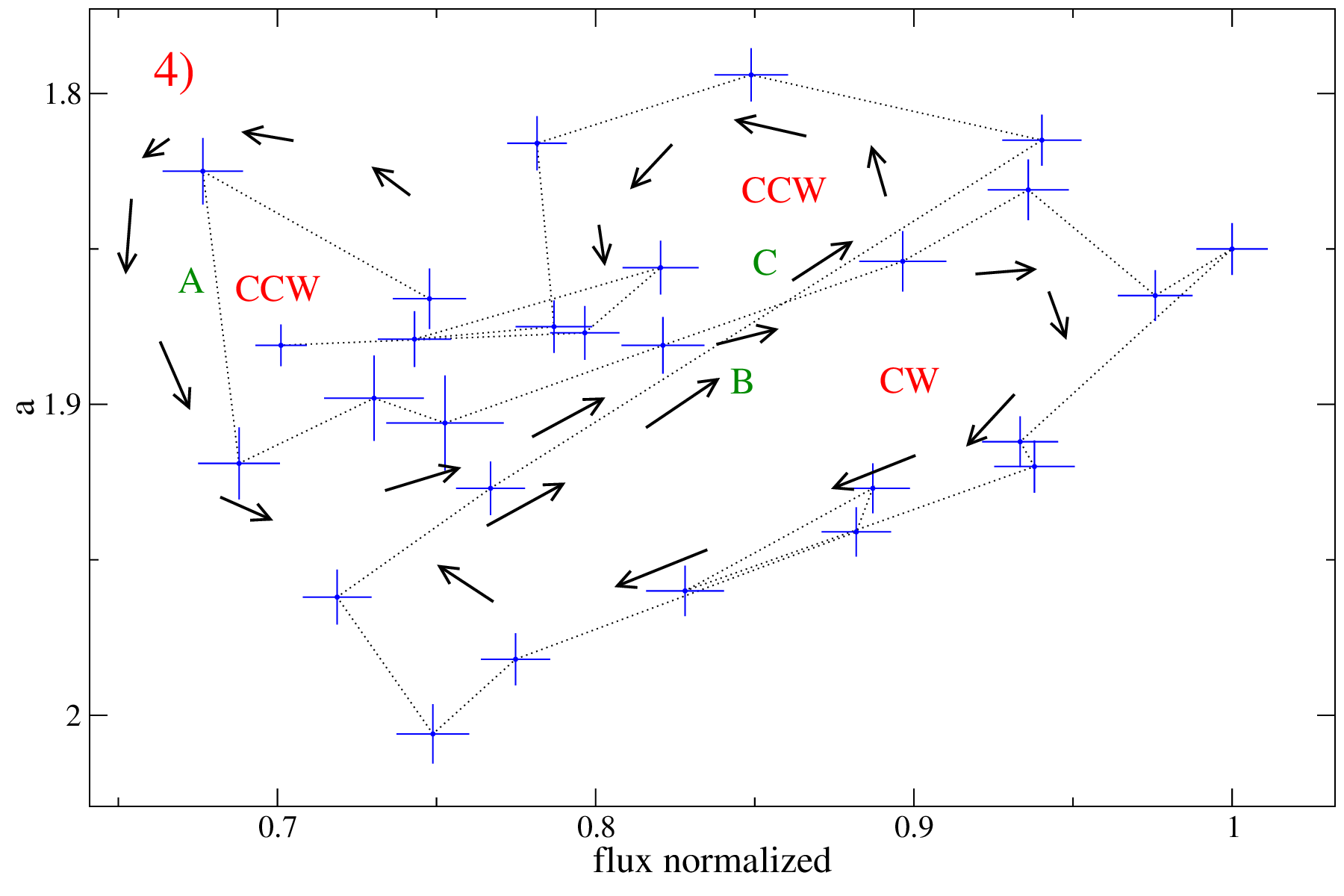,width=5.5cm,angle=-90}&
\epsfig{file=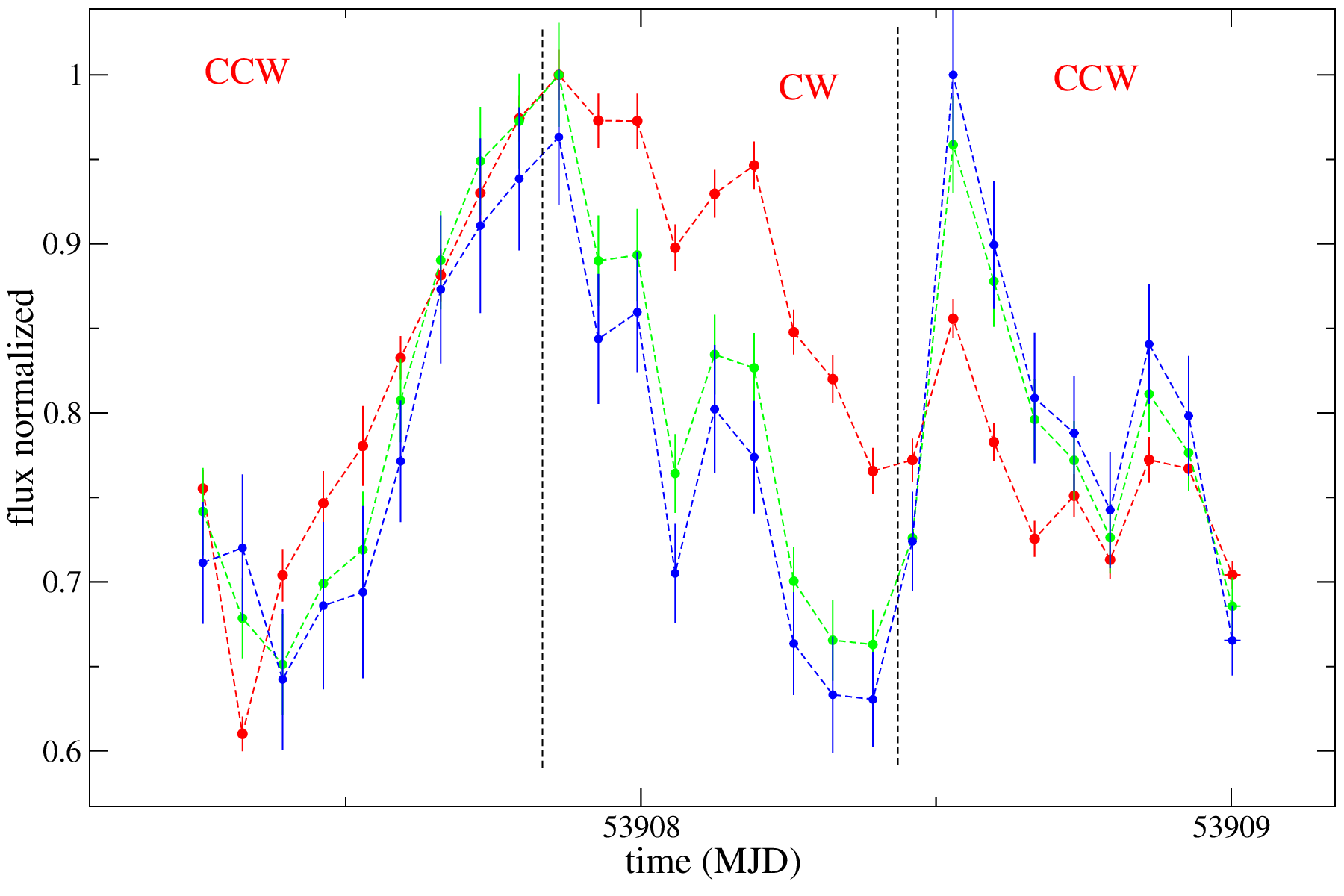,width=5.5cm,angle=-90}\\
\end{tabular}
\caption{
\textit{Left} panels: Spectral patterns in the $a$-flux plane showing clockwise and counterclockwise trends.~
\textit{Right} panels: \swx~ light curves at three different bands: soft ($0.2-3.0$ keV, red), medium ($3.0-5.0$
keV, green), and hard ($5.0-10.0$ keV, red) . Capital letters guide the reader during the comment in the
text. In both left and right panels, the fluxes are normalized to the maximum value reached during that particular time-interval.
}
\label{fig-patt-1}
\end{center}
\end{figure*}

\begin{figure*}[]
\begin{center}
\begin{tabular}{ll}
\epsfig{file=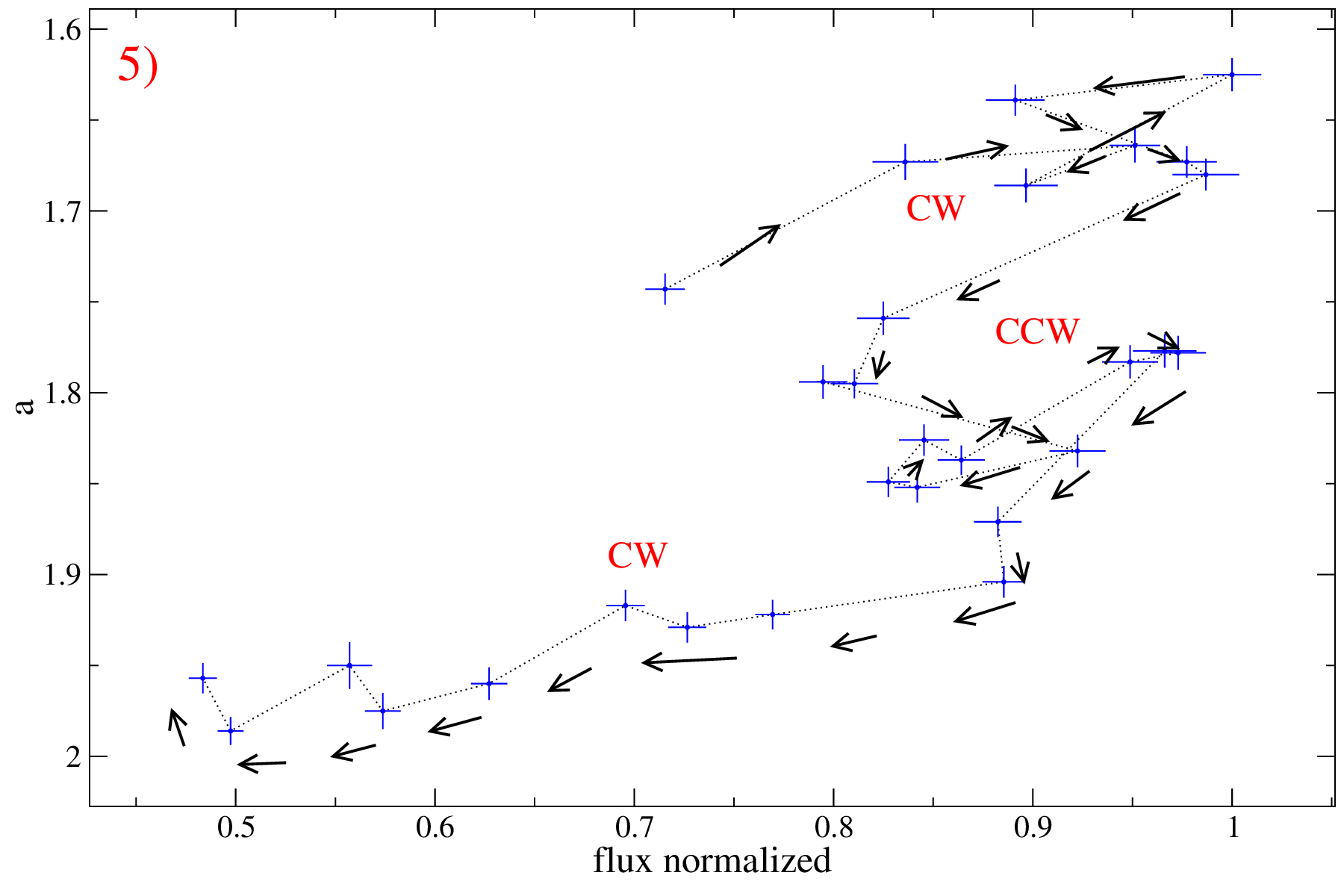,width=5.5cm,angle=-90}&
\epsfig{file=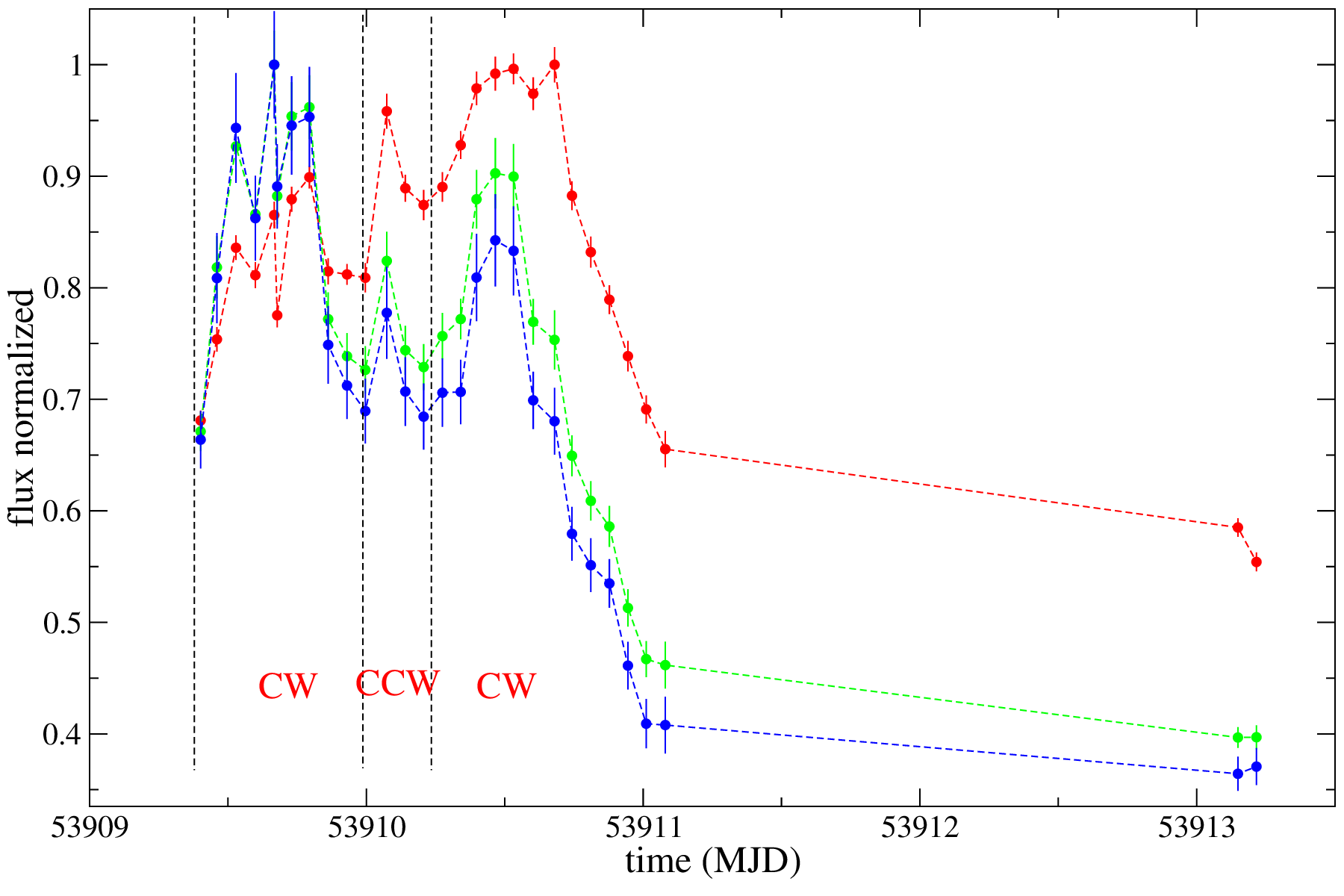,width=5.5cm,angle=-90}\\
\end{tabular}
\caption{
\textit{Left} panels: Spectral patterns in the $a$-flux plane showing clockwise and counterclockwise trends.~
\textit{Right} panels: \swx~ light curves at three different bands: soft ($0.2-3.0$ keV, red), medium ($3.0-5.0$
keV, green), and hard ($5.0-10.0$ keV, red) . Capital letters are referenced in the discussion in the
text. In both left and right panels, the fluxes are normalized to the maximum value reached during that particular time interval.
}\label{fig-patt-2}
\end{center}
\end{figure*}

In this section  we follow the classical approach of the hardness ratio (HR)
analysis, to investigate  cooling and/or  acceleration features.
The HR is evaluated as the ratio
of the $2.0-10.0$ keV band to the $0.2-2.0$ keV band.\\	 
Analysis of the evolution of the HR shows that the
correlations between variations in the soft and hard bands results in
modulation of the HRs, with the spectra harder when the source is brighter and
softer when weaker. This trend is clearly  visible in Fig. \ref{fig-HR-flux}. 
A significant scatter in the points hints that the dynamic among different flares 
is quite different due to different underlying physical conditions.\\
Moreover, we note  that all but 3 of the points in the shaded area belong 
to the period from 06/23/2006 to 06/27/2006 and to the 07/15/2006 pointing, 
when the source was  brighter than in the previous pointings.\\ 

A deeper understanding of the spectral dynamics can be achieved from looking at 
the hysteresis patterns of the single flares in the $a$-flux or $HR$-flux plane, 
where $a$ is the photon index  at $1$ keV (see Tab. \ref{tab-fit}).\\
The time scales relevant to understanding the patterns in the $a$-flux plane are:
the injection ($\tau_{inj}$), the escape ($\tau_{esc}$), the cooling ($\tau_{cool}$),  the
acceleration  ($\tau_{acc}$) and the light crossing ($\tau_{cross}$) time.
According to \citet{Kirk1998}, the loops are
expected to be clockwise (CW) and with soft lag when the flare is observed at 
frequencies where the higher energy variability is faster than at lower energy. 
(as in the case of synchrotron cooling).
On the contrary, when observed at frequencies for which the acceleration and cooling 
time scale are almost equal, the loops are expected to be counter-clock-wise (CCW) 
with a possible  hard lag.\\
We investigated carefully all the patterns from our data-set, 
and in Fig. \ref{fig-patt-1} and Fig. \ref{fig-patt-2} we report the 5 flares 
showing a clear CW or CCW loop. 
The left panels show the evolution of $a$ as a function of the flux, and the 
right panels show the light curves in three different bands. The fluxes are 
normalized to the maximum value reached in that particular time interval.
As a general result, we did not find any significant plateau among the light 
curves of these flares, meaning that  $\tau_{inj}<\tau_{cross}$. We describe the 
behaviour of the flares individually:
\begin{description}

 \item{Flare 1)} The flare has a CCW pattern; it begins with a softening of the 
spectrum with a   flux almost steady (A) probably reflecting the cooling from 
the previous flaring episode, followed by a flux increase with a spectral 
hardening (B). 

 \item{Flare 2)} The flare has two patterns: one CW and one CCW. It starts with 
a decreasing flux and a spectral softening (A) ( $\tau_{cool} \ll \tau_{esc}$). 
Figure \ref{fig-patt-1}  (left panel) shows that the flux in the soft
band is still increasing when the fluxes in the medium and hard bands are
already decreasing.
When the spectrum starts to get harder (B) the flux is still decreasing, this may 
hint that we are seeing the propagation of the new injection starting from the 
hard band, with the soft band still decreasing.  Then, the flux starts increasing 
with the spectrum hardening (C). When the flare starts to decay (D) the pattern 
switches to CCW and we  see a decrease of the flux with a spectral softening.

 \item{Flare 3)} The flare has a CCW pattern. starts with a flux 
decrease and a spectral softening (A), then the flux increases and the 
spectrum hardens (B). In the last part (C), the flux decreases and the spectrum 
still hardens. As in the previous case this may hint the start of a new
hard flaring component.	

 \item{Flare 4)} This flare has three patterns: CCW,CW, and CCW.  It starts with 
a CCW loop. Initially there is a spectral change  with an almost steady flux (A), 
followed by a weak spectral hardening with a flux increase (B). The CW loop 
is dominated by  the cooling time. The final CCW pattern is characterized 
by a flux increase and a very rapid spectral hardening. 
Also in this case the new flaring component seems to start from the hard band. 
The  flux decrease (C)  is almost achromatic in  this case, and the escape time 
may be dominant over the cooling one ( $\tau_{esc} \ll \tau_{cool}$).

\item{Flare 5)} This flare has  three patterns: CW,CCW, and CW.  It does not have 
features different form the previous ones.
\end{description}

CW loops were observed for the same source in the past by \citet{Taka1996} with \asca~ observations.
\citet{Zhang2002} observed  CCW in \sax~ observations on 21 April 1998. Similar kind of
loops were also observed in PKS2155-304 by \citet{kata2000} and 1H1426+428 by \citet{Falcone2004}.\\



\section{Spectral parameter trends.}
We compare  the trends among the spectral parameters with those resulting from 
the statistical analysis in \citet{Trama2007b}. We extend the data set 
presented in that work with the results from the observations analysed in this 
paper. 
It is worth noting that the whole data set  spans about ten years, sampling 
the source in very different brightness states, making the final result more 
statistically significant.\\



\subsection{$E_p-b$ trend}
\begin{figure}[t]
\begin{center}
\epsfig{file=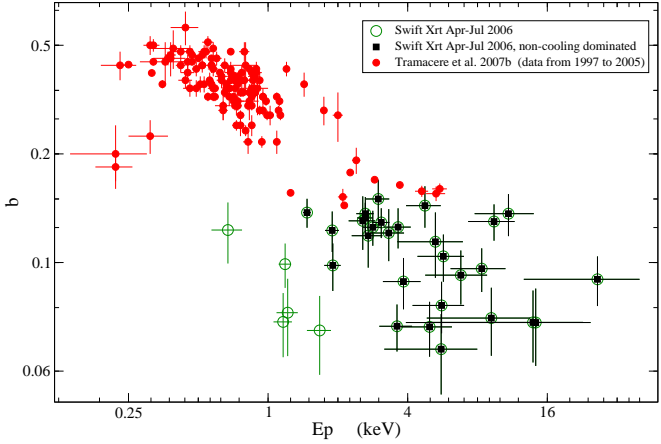,width=6cm,angle=-90}
\caption{Scatter plot of  the curvature ~($b$) vs. $E_p$. Red circles represent data from
\citet{Trama2007b} spanning from 1997 to 2005, instruments: \asca, \sax, \xmm. Black boxes represent
\swf~ data from the present analysis without the cooling-dominated events. Empty green circles
represent the whole \xrt~ data set presented in this paper.}
\label{fig-Ep-b}
\end{center}
\end{figure}

The evolution of  the curvature parameters as  a function of the peak energy  
points out relevant features of the acceleration process. 
This analysis
indicates that as the peak energy of the emission increases, the cooling time
scale shortens and can compete with the time scales for acceleration.
In this case, it is possible to observe a  bias in the $E_p-b$ relation, 
due to the cooling time scale dominating over time scales for acceleration.\\ 
We analysed the data in Tab. \ref{tab-fit-1}, identifying cooling-dominated 
observations (lines labeled with (c)), characterized by a strong flux decrease 
and strong spectral softening ($a\simeq2$). In Fig.  \ref{fig-Ep-b} we plot with 
green empty circles the whole \xrt~ data set from Tab.  \ref{tab-fit-1} and with 
black squares the points without the strong cooling contamination
discussed above.\\
In the same figure we report also data from \citet{Trama2007b}, showing that
 the trend in our sample is consistent with that from the historical data. 
This agreement  confirms that the  spectral curvature  decreases as $E_p$ moves 
toward higher energies. \\

This phenomenon can be interpreted according to two different scenarios.
A first scenario is that of an  energy-dependent acceleration probability process 
(EDAP). Within this context, \citet{Massaro2004a} showed that for acceleration 
efficiency  inversely proportional to the energy itself, the energy distribution 
approaches a log-parabolic shape.  According to this model, the curvature ($r$) 
is related to the fractional acceleration gain ($\varepsilon$)  
by $r\propto \frac{1}{\log \, \varepsilon}$. A possible example  is given by 
particles confined by a magnetic field, whose confinement efficiency ($P_{acc}$) 
decreases as the gyration radius ($R_L$) increases. From $E_p\propto \varepsilon$ 
and $ r\propto \frac{1}{\log \, \varepsilon}$  the negative trend 
between $E_p$ and $b$ follows. This trend is  in agreement with the observed trend.\\

An alternative explanation is provided by the stochastic acceleration framework (SA), 
with the presence of a momentum-diffusion term. In this scenario, the diffusion 
term acts on the electron spectral shape broadening the distribution. 
In particular,  \cite{Karda1962} showed that a log-parabolic spectrum results 
from a Fokker-Planck equation with a momentum-diffusion term and a mono-energetic 
or quasi-mono energetic particle injection. 
The results presented in \citet{Trama2007b} rely on the theoretical prediction 
from the  \cite{Karda1962} model, that the curvature term  ($r$) is inversely 
proportional  to the diffusion term ($D$):
\begin{equation}
 \label{eq-r}
  r \propto\frac{1}{Dt}~~.
\end{equation}
This relation  leads to the following trend  among the  peak energy of the 
electron distribution
($\gamma_p$), the synchrotron curvature ($b=r/5$), $r$, and $E_p$:
\begin{equation}
\label{eq-Ep}
 \ln(E_p)= 2\ln(\gamma_p)+3/(5b).
\end{equation}
In paper II \citep{Tramacere2009} we will discuss in detail how the stochastic 
acceleration  can be used to reproduce this trend drawing a physical scenario 
that fits this phenomenological picture. Here, we just remark that both the 
momentum-diffusion term $D$ and, the fractional acceleration gain $\varepsilon$ 
can explain the anticorrelation observed between $E_p$ and $b$.


\begin{figure}[t]
\begin{center}
\epsfig{file=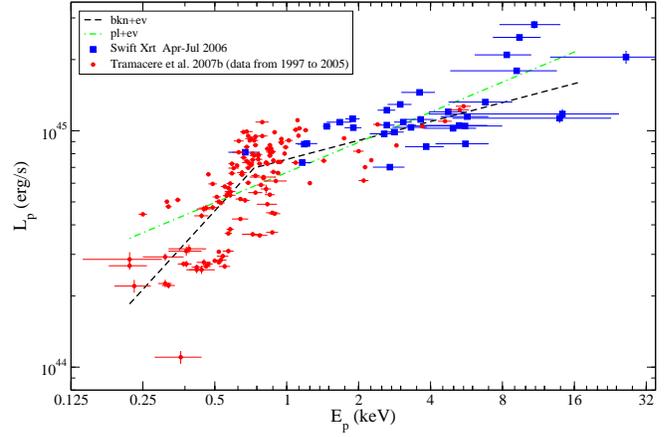,width=5.8cm,angle=-90}
\caption{Scatter plot of $L_p$ vs $E_p$. Red circles represent data from \citet{Trama2007b} spanning
from 1997 to 2005, instruments: \asca, \sax, \xmm. Blue boxes represent \swf~ data from the present
analysis. The dashed dotted line represents the best fit using a power-law  and the dashed line
represents the best fit using a broken power law.}
\label{fig-Sp-Ep-fit}
\end{center}
\end{figure} 

\subsection{$S_p$($L_p$)-$E_p$ trend}
The trend between $S_p$ and $E_p$ provides interesting indications concerning 
the driver of the spectral changes in the X-rays, in terms of the synchrotron 
emission. This trend can be used to understand how the luminosity of the 
jet evolves  as the particle energy increases.
In the framework of the synchrotron theory \citep{Ryb1979}, the dependence of $S_p$
on $E_p$ can be expressed in the form of a power-law:
\begin{equation}
 S_p \propto E_p^{\alpha}.
\end{equation}

In fact, starting from the following functional relation the SED peak height reads: 

\begin{equation}
  S_p \propto n(\gamma_{3p})~\gamma_{3p}^{3}~B^2~\delta^4 
\end{equation}

and the peak energy is given by:

\begin{equation}
 E_p \propto \gamma_{3p}^2B~\delta
\end{equation}

where $\gamma_{3p}$ is the peak of $n(\gamma)~\gamma^{3}$, $B$ the magnetic field, 
and  $\delta$ is the beaming factor.\\
If we take into account a log-parabolic distribution for the electrons emitting 
the X-ray photons, then it is easy to show that the total emitters number 
$N=\int n(\gamma)d\gamma \simeq n(\gamma_p)~\gamma_p$, with $\gamma_{p}$ the 
peak of $n(\gamma)~\gamma$. This in terms of $S_p$ implies

\begin{equation}
 S_p \propto N~\gamma_p^{2}~B^2~\delta^4. \\
\end{equation}

Thus $\alpha=1.0$ applies when the spectral changes are dominated only by 
variations of the electron average energy and $N$ is constant. $\alpha=1.5$ 
applies when the spectral changes are dominated by variations of the average 
electron energy but $N$ is not constant; $\alpha=2$ as for changes of the magnetic 
field; $\alpha=4$  if  changes in  the beaming factor dominate; formally, 
$\alpha=\infty$  applies for changes only in the number of emitting particles, 
which implies either a change in the electron density  or a change in the source 
size. The case $\alpha=1.0$  holds only if the electron distribution is  a 
log-parabola and if the curvature is constant.\\
In order to have a deeper understanding of the relation to the jet energetics, 
we plot on the $y$  axis of Fig. \ref{fig-Sp-Ep-fit}, $L_p= S_p~4\pi~ D_L^2$, where 
$D_L\simeq 134.1$ Mpc is the luminosity distance\footnote{We used a flat 
cosmology model with $H_0=0.71$ km/(s Mpc) $\Omega_M=0.27$ and 
$\Omega_{\Lambda}=0.73$.}. We report the $L_p$ values both in Tab. \ref{tab-fit} 
and \ref{tab-fit-1}.
The resulting scatter-plot represents the $L_p-E_p$ trend obtained from merging our 
data sample with that from \citet{Trama2007b}.\\ We fitted the data by means of 
a simple power law (PL) $L_p \propto E_p^{\alpha}$, and a broken power-law (BPL)
 
\begin{eqnarray}
 L_p &\propto &E_p^{\alpha_1},~~~~  E_p \leq E_b \\
 L_p &\propto &E_p^{\alpha_2},~~~~  E_p > E_b. \nonumber
\end{eqnarray}

The simple power-law fit gives a value of $\alpha=0.42\pm 0.06$, and the
broken power law gives a break energy $E_b=1\pm 1$ keV, with the two slopes 
$\alpha_1=1.1\pm 0.2$ and  $\alpha_2=0.27\pm 0.07$.\\
We focus on the results from the BPL fit. Although the break energy is not well 
constrained, it is worth to note that spectral slopes are quite different with a 
high statistical significance. This break in the trend implies that for 
$E_p\lesssim$ 1keV and $L_p\lesssim10^{45}$ erg/s the driver follows the 
relation with $\alpha \simeq 1.0$ (we define this state the \textit{quiescent} sate), 
whilst for $E_p \gtrsim$ 1keV and $L_p \gtrsim 10^{45}$ erg/s, the driver is 
ruled by $\alpha \simeq 0.2$ (we define this as the $high$ state).\\
In Paper II \citep{Tramacere2009} we will discuss in detail how the stochastic 
acceleration or the energetics of the jet can be used to reproduce this trend.



\begin{figure}[h]
\epsfig{file=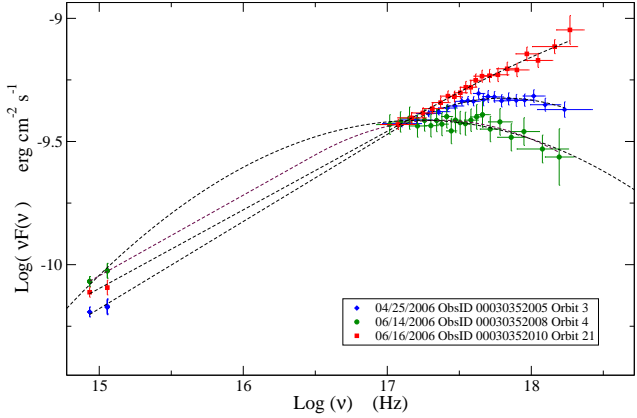,width=6.2cm,angle=-90}
\caption{Three different spectral shapes from the data presented in this
paper. Red boxes represent a power-law spectrum  observed during Orbit 3 (ObsID 00030352010) on 06/16/2006. Blue
diamonds represent a spectrum that is a log-parabola with a low energy power-law tail, 
from Orbit 3 (ObsID 00030352005) on  04/25/2006. Green circles
represent a log-parabolic spectrum from Orbit 4 (ObsID 00030352008) on 06/14/2006.}
\label{fig-SED-UVOT-XRT}
\end{figure}

\section{The \swu~ \swx~ connection: the low-energy power-law tail in the electron 
distribution.} Both $E_p-b$ and $S_p-E_p$ trends allowed us to understand 
the shape of the electron distribution for particles emitting at energies close 
to $E_p$. Anyway, the particle energy distribution   can develop a low energy 
shape  quite different from the extrapolation of the high energy branch. This 
difference is relevant for discriminating among different acceleration 
processes.
In this perspective, the connection  between the UV and X-ray spectra  can 
give useful information about the low energy tail of the electrons emitting 
in the X-ray. We analysed carefully all of the spectra with simultaneous X-ray and 
UV observations. As result we found that joint \uvt-\xrt~ SEDs can be classified 
in three categories:
 \begin{description}
 \item[a)] described by a log-parabola (LP) 
 \item[b)] described by a power law (PL)
 \item[c)] described  by a spectral law that is power law in the low energy tail, 
turning into a log-parabola in the high energy one (LPPL) \citep{Massaro2006}, 
whose functional form can be expressed by
\begin{eqnarray}
\label{lppl-sed}
 \nu~F(\nu)&=&N~(\nu/\nu_c)^{-a_{\nu}},~~~~~~~~~~~~~~~~~ \nu \leq \nu_c \nonumber  \\
 \nu~F(\nu)&=&N~(\nu/\nu_c)^{-(a_{\nu}+b~\log(\nu/\nu_c))} , \nu >\nu_c~~~.
\end{eqnarray}
where $a_{\nu}$ is the spectral index of the SED ($\nu~F(\nu)$), and $\nu_c$ is 
the frequency where happens the turn-over in the SED ( Fig. \ref{fig-SED-UVOT-XRT}).\\
\end{description}

From the analysis of this  spectral behaviour it is possible to constrain  the 
minimum energy of the radiating electrons. In fact electrons radiating mainly in 
the UV band have a Lorentz factor $\gamma_{UV}$ satisfying the following 
condition  \citep{Ryb1979}:
\begin{eqnarray}
\label{sync-freq}
 10^{15}Hz  &\simeq& 3.7\times10^{6}~B~\delta~{\gamma_{UV}}^2/(1+z)  \\
\gamma_{UV} &\simeq& 1.6 \times10^{4} \sqrt{\frac{1+z}{B~\delta}} \nonumber.
\end{eqnarray}
If the spectral shape is consistent with the same log-parabola extending from 
the X-ray band  down to the UV band (case a), then it means that electrons 
radiating at UV frequencies belong to the same electron population and according 
to Eq. \ref{sync-freq} we have $\gamma_{min} \lesssim \gamma_{UV}$.\\
The condition $\gamma_{min} > \gamma_{UV} $ may occur when we observe a 
PL (case b) or a LPPL  (case c). 
In fact, if $\gamma_{min} > \gamma_{UV} $ then the spectra in the X-ray-to-UV band 
will be described by the asymptotic low-energy approximation of the single particle 
synchrotron emission that is a power law with slope 
$a_{\nu}\simeq{-4/3}$ ($SED~\propto\nu^{4/3}$) \citep{Ryb1979}.\\
For both   case b) and c) we note that our data  give $a_{\nu}\simeq[0.25-0.4]$,  
a value very different from the asymptotic synchrotron kernel expectation. 
This hints that both cases the UV photons are  likely to originate from 
an electron  distribution that has a power law tail in the energetic range 
radiating in the  UV-to-soft-X-ray band. A phenomenological option to explain the 
case c) is  an electron distribution  that is a power 	law at low energies with a 
log-parabolic high-energy branch (LPPL) \citep{Massaro2006}:
\begin{eqnarray}	
\label{lppl-elec}
 n(\gamma)&=&K~(\gamma/\gamma_c)^{-s},~~~~~~~~~~~~~~~~~ \gamma \leq \gamma_c \nonumber  \\
 n(\gamma)&=&K~(\gamma/\gamma_c)^{-(s+r~Log(\gamma/\gamma_c))} , \gamma >\gamma_c~~ ,
\end{eqnarray}
where $\gamma_c$  is the turn-over energy.\\
In the case b), the electron distribution is assumed to be a pure power-law.\\ 
Interestingly  for case b) and case c) we can also constrain  the typical slope 
of the power-law branch of the electron distribution, using the well known relation
between the spectral index in the particle distribution $s$ and that in the SED 
\citep{Ryb1979}:
\begin{equation}
SED \propto \nu^{-a_{\nu}}=\nu^{-(s-3)/2}.
\end{equation}
For the typical values of $a_{\nu}$ observed in our data set, the resulting value 
of $s$ is in the range $s\simeq[2.2-2.5]$.\\

The presence of  a power-law feature and the range of observed  spectral 
indices   are relevant both in the context of Fermi first-order acceleration
models and from an observational point of view.\\
From the observational side , it is worth to remark that  \citet{Waxman1997} and 
\citet{Mesz2002}, studying the  the afterglow X-ray emission of $\gamma-$ray 
bursts (GRB), inferred an electron distribution index  $s\simeq 2.3 \pm 0.1$.  
This is very close to those found in our data,  but coming from a quite different 
class of sources.

From a theoretical point of view,  there are several works concerning relativistic
shock acceleration models that start from different analytical or numerical 
approaches and find values of $ s \simeq [2.2-2.4]$ \citep{Achterberg2001,Gallant1999,
Lemoine2003,Blasi2005,Ellison2004}.
These values are  consistent with those from our data set.\\  
The power-law feature  is also consistent with a purely stochastic scenario.
Anyway, the usual limitation of the stochastic model to explain a  
\textit{universal} index   relies on the fine tuning required on the ratio of 
the acceleration time scale to the loss time ($s\simeq 1+t_{acc}/t_{esc}$),   
in order to match the observed values.\\
Moreover, we stress that a power-law electron distribution 
$n(\gamma) \propto \gamma^{-2.3}$  is not compatible with a Maxwellian-like 
distribution  ($n(\gamma) \propto \gamma^{2}$) resulting from  the equilibrium 
of SA processes without relevant particle escape.\\

In conclusion both  case c) and b)  are better explained better by a first 
order process. We will discuss this topic further in Sec. 9.\\

\begin{figure*}[]
\begin{center}
\begin{tabular}{ll}
\epsfig{file=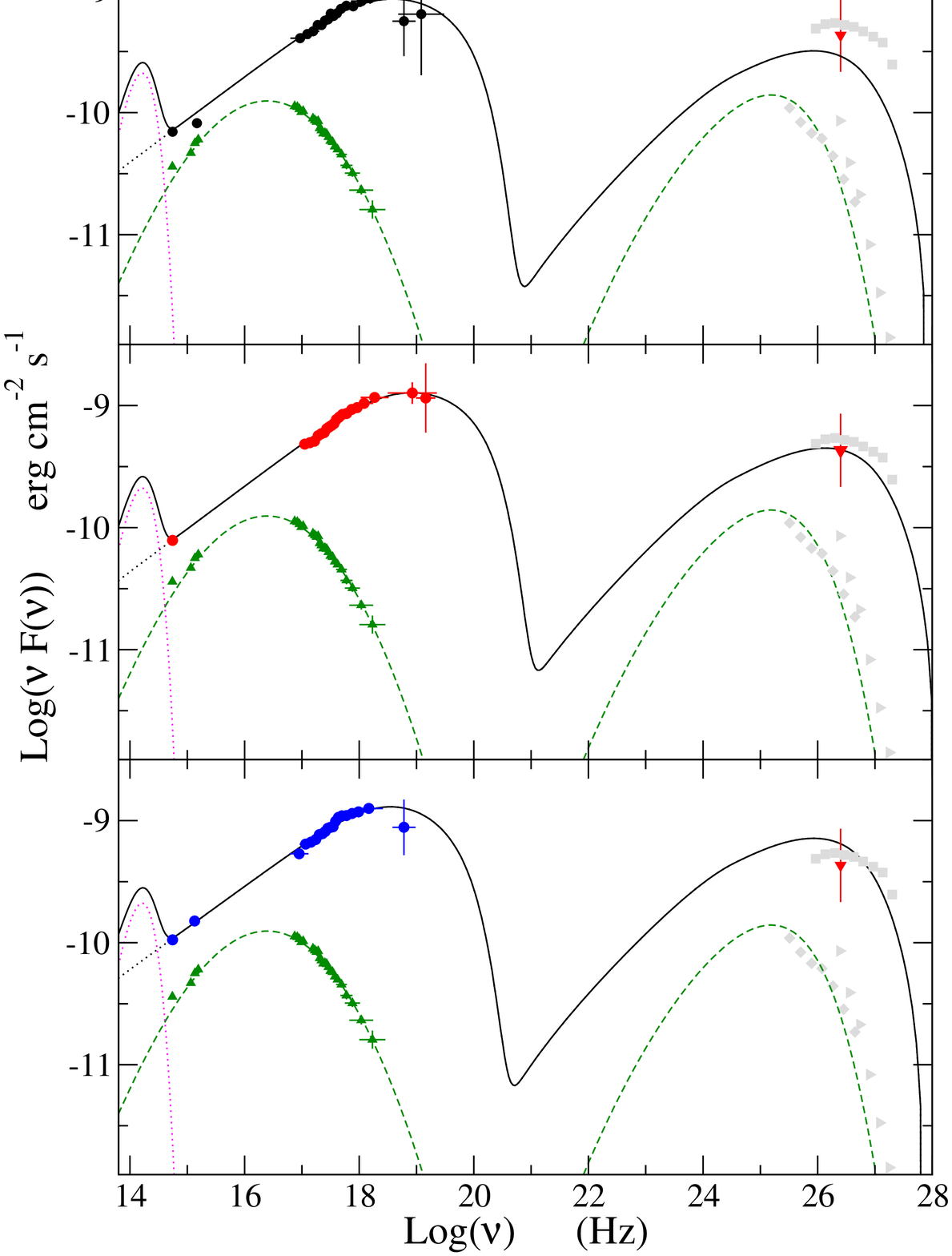,width=8.95cm,angle=0}
&\epsfig{file=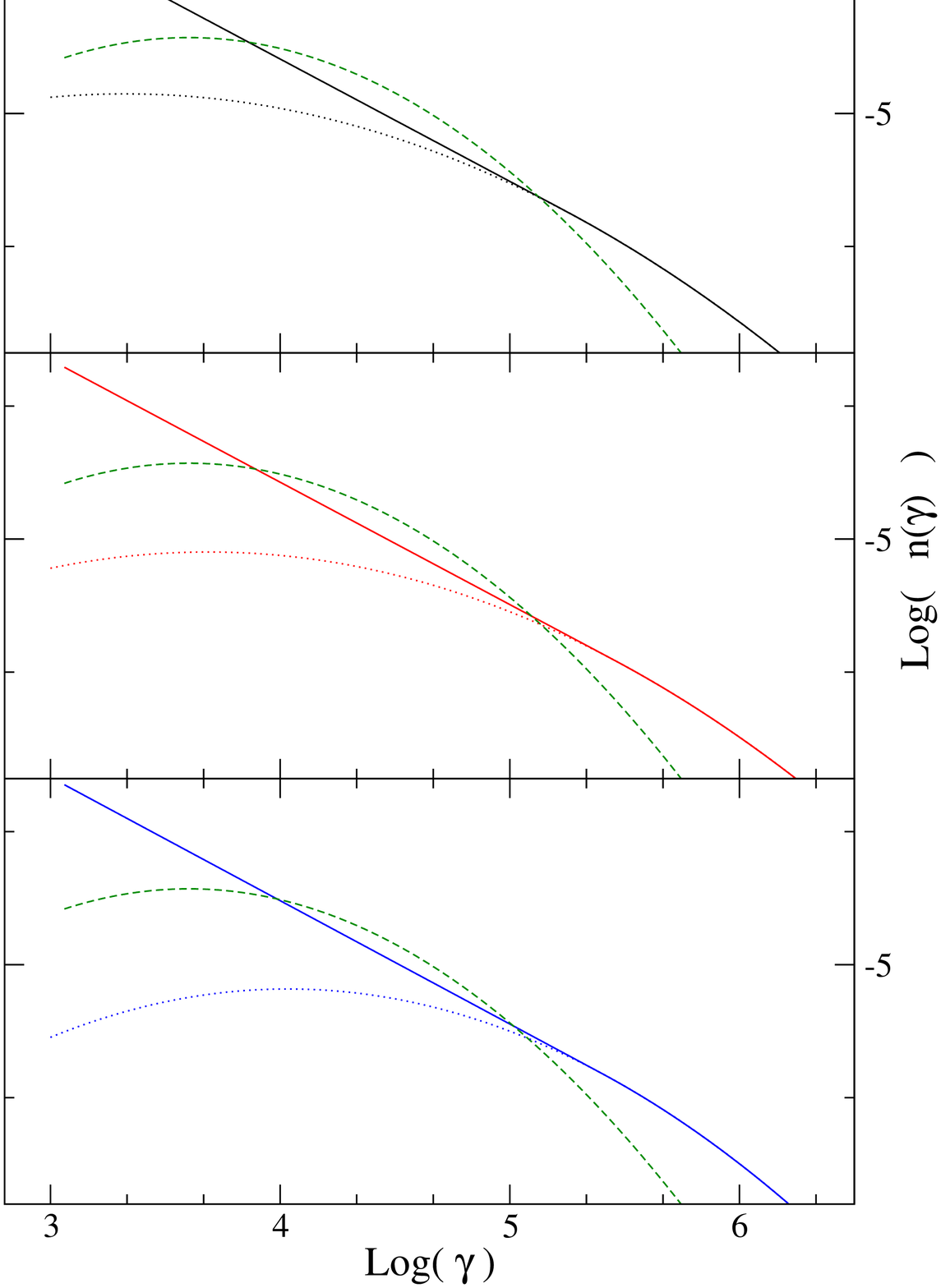,width=8.95cm,angle=0}\\
\end{tabular}
\end{center}
\caption{SSC fits of three different observations with simultaneous \uvt~ \xrt~ and \bat~ data.
\textit{\textit{Left}} panels show the SSC model,  from top to bottom:
solid circles represent data from  04/22/2006, 06/23/2006, 07/15/2006 \swf~ observations. Green  
triangles show  \swx~ data on 03/31/2005, data from \citet{Trama2007b}. Solid grey polygons
represent non-simultaneous EBL corrected TeV data. Solid gray squares represent the high state on 2001
observed by \whipple, data from \citet{Albert2007}. Solid gray diamonds represent the average
2004-2005 TeV spectrum as observed  by \magic~ \citep{Albert2007}. Solid gray right triangles represent
the average spectrum from December 2005 to February 2006 as observed by \tactic~ \citep{Yadav2007}.
The solid red triangle represents a \whipple~ observation on June 18,19,21 from \citet{Lich2008},
that is very close in time to our 06/23/2006 data set. The solid lines represent the best fit by a
SSC model to our simultaneous \swf~ observation, and the dashed line is the best SSC fit to the
\swf~ data on 03/31/2005. The dotted lines represent the modeling of the galaxy contribution by means of
black body spectral shape.
 \textit{\textit{Right}} panels show the electron  distributions for the SSC models in the left panels.
The solid lines  in the right panels represent the electron distributions for the best fit models of the
three 2006 \swf~ observations, the dotted lines  represent the extrapolation of the LP branch of the LPPL
distribution, and the dashed lines represent the electron distribution for the 03/31/2005 data.
}\label{fig-FIT-SED-UVOT-XRT}
\end{figure*}

\section{SED modeling and GeV/MeV predictions.}

We model  the SEDs of the three observations that have simultaneous
\xrt,~\bat~ and \uvt~ data, using  a standard one-zone SSC scenario. 
The only useful TeV data   found in the literature are   
from \whipple~ observations  on June 18,19, and 21  \citep{Lich2008}. These data,
almost simultaneous only with the   06/23/2006 \swf~ pointing,
provide only the TeV flux, without giving  a description of the spectrum. For
this reason   they are used only to estimate  the  TeV flux level during that 
pointing.

To have a feeling of the spectral and flux range of variability, we also 
plot \swx~ data from  03/31/2005 \citep{Trama2007a}, and some TeV SEDs representing 
the source in different flaring states \citep{Albert2007,Yadav2007} 
(see left panel of Fig. \ref{fig-FIT-SED-UVOT-XRT}).\\

The 2006 SEDs that we want to model  have a power-law spectral dependence between 
the \uvt~ and \xrt~ bands. As described  in Sect. 7, the most generic distribution 
accounting for this  spectral shape is a power-law at low energy with a log-parabolic 
high-energy branch (Eq. \ref{lppl-elec}). 
On the contrary, the 03/31/2005 SED can be modeled using a log-parabolic electron 
distribution that we  express in terms of the peak energy as (LPEP):
\begin{equation}
 n(\gamma)= K~10^{-r~(\log(\gamma/\gamma_p))^2}.
\label{eq-lpep-elec}
\end{equation}

Since we do not know the actual shape and flux at TeV energies (we have only an 
estimate of the flux at $\simeq 1$ TeV) we cannot constrain in detail the SSC 
model using the canonical $B-\delta$ plane analysis \citep{Tavecchio1998}. 
Yet, we can still  use the results reported in Sect. 3 and 7 to constrain
 some of the SSC parameters.\\
The first constraint comes from the source variability (see Sect. 3):
\begin{equation}
 R/\delta \leq 1 \times 10^{14} ~\mbox{cm}.
\end{equation}
A second constraint comes from the analysis of the \uvt~ connection with the 
\xrt~ data that gives an upper limit to $\gamma_{min}$
\begin{equation}
\gamma_{min} \lesssim 1.6 \times10^{4} \sqrt{\frac{1+z}{B~\delta}}. 
\end{equation}
To constrain  the value of the maximum electron energy we can use the 
maximum energy of the synchrotron emission,  taking as the corresponding energy the 
most energetic bin of the \bat~ detector with a significant signal  ($\simeq$ 50 keV)
\begin{equation}
\gamma_{max} \gtrsim 1.8 \times10^{6} \sqrt{\frac{1+z}{B~\delta}} 
\label{eq-gmax-kev}.
\end{equation}
A  further constraint on $\gamma_{max}$ can be set by estimating the typical electron
energy required to produce TeV photons \citep{Bednarek1999}.
Based on the historical data for  \mrk, the spectrum typically reaches up to 10 
TeV, and this implies

\begin{equation}
\gamma_{max} \gtrsim   2\times 10^7 \frac{1+z}{\delta}~~~.
\label{eq-gmax-tev}
\end{equation}
And following \citet{Bednarek1999}, combining Eq. \ref{eq-gmax-tev} and 
\ref{eq-gmax-kev} we obtain an upper limit on the magnetic field:
\begin{equation}
\label{eq-Bdelta}
B/\delta \lesssim 0.008 ~ \textnormal G. \\
\end{equation}

The best fit model was obtained by combining our numerical SSC code \citep{Trama2007PhD} 
with a numerical minimizer. According to the observationally derived constraints, 
we fixed the value of the beaming factor at $\delta=25$, the magnetic field $B=0.1~G$,
the source size $R=2.1\times10^{15}$ cm, and tuned $\gamma_{min}=1100$ to get 
the right Compton  dominance, leaving $r,~s,~\gamma_c$, $\gamma_{max}$ and $N$ 
as free parameters.  The resulting best fit parameters are reported in Tab. 
\ref{ssc-fit} and \ref{ssc-fit-1}.\\ 

In the left panels of Fig. \ref{fig-FIT-SED-UVOT-XRT} we show the best fit results 
for the SSC model for the 2006 data and for the 31/03/2005 pointing. In the right 
panels we show the corresponding electron distributions.
The values of the electron curvature are consistent with those observed in the 
X-ray emission according to the relation $b\simeq r/5$.
The typical value of the ratio $u_e/u_B \simeq [120-180]$ results in a particle 
dominated jet in agreement with the typical values for TeV HBLs peaking in the 
hard X-ray \citep{Kino2002,Sato2008}.
The low-energy power law branch of the electron energy distribution 
follows $n(\gamma) \propto \gamma^{-2.3}$, in agreement with the analysis 
presented in Sect. 7.    

We remark that UV data are weakly contaminated by the host galaxy emission.

Regarding this, we plot in the left panel of Fig. \ref{fig-FIT-SED-UVOT-XRT} 
the contribution of the galaxy emission modeled as a black body spectral shape 
(dotted-line). This plot clearly shows how the \uvt~ data taken into account 
lie beyond the galaxy emission spectral  cut-off.
On the contrary, during the low states, the possible power-law tail would lay 
typically at optical and lower frequencies, where the emission  is typically
dominated by the galaxy. This implies that we actually don't know whether 
the electron distribution has such a power-law tail also during the lower
state. The SED model of data on 31/03/2005  shows in fact that the \uvt~ 
points are compatible with the synchrotron emission from a log-parabolic  
electron population. \\
Moreover, a  power-law low-energy branch if present in the 2005 SED would 
yield an index  harder than those from 2006, but with much lower flux.

\begin{figure}[!h]
\epsfig{file=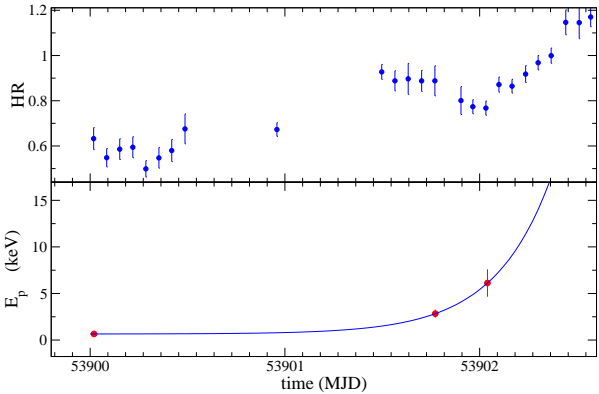 ,width=5.8cm,angle=-90}
\caption{\textit{Lower} panel: the increase of $E_p$ during a flare starting on 06/14/2006, fitted
by an exponential function. \textit{Upper} panel: the corresponding HR.}\label{fig-Ep_acc_rate}
\end{figure}

Another interesting analysis can be performed by looking at the acceleration time 
scales determined by the rate of the variation of $E_P$. In particular, we can use 
the $e$-folding time of  $E_p$ to infer the electron acceleration time scale  
$\tau_{acc}$. In detail,  from $E\propto \gamma^2$ it follows that the rate of 
change of $E_p$ can be linked to the rate of change of $\gamma$ for electrons 
radiating in the hard X-ray band. In Fig \ref{fig-Ep_acc_rate} we show a
flare having well constrained values of $E_p$  for the rising side. In this 
figure we plot the values of $E_p$ as a function of  time. The $e$-folding 
time for  $E_p$  results in a $\gamma$ $e$-folding time of about 0.15 days, 
which  is the value of the typical acceleration time scale. This is the time scale 
that will compete with the cooling one to generate the equilibrium in the
$n(\gamma)$ distribution. In particular taking into account the synchrotron cooling:
\begin{equation}
\tau_{cool}\simeq 10^8/(B^2\delta\gamma) ~~ s
\end{equation}

the equilibrium of the distribution for  a log-parabolic or Maxwellian-like 
distribution is the peak of $n(\gamma)$. In the case of LPPL distribution it 
will be constrained between $\gamma_{c}$ and $\gamma^{LP}_p$. Imposing 
$\tau_{cool}=\tau_{acc}$ we find:  
\begin{equation}
 \gamma_{eq}=\frac{10^8}{\tau_{acc}~B^2\delta}
\end{equation}
For $\tau_{acc}\simeq 0.15$ days,  $\delta=25$ and using   $\gamma_{eq}$ of the 
order of $10^4$ the resulting value of the magnetic field is 
\begin{equation}
B=\sqrt{\frac{10^8}{\tau_{acc}\delta\gamma_p}}\simeq 0.16~ G
\end{equation}
which is consistent with the upper limit of Eq. \ref{eq-Bdelta} and with our best 
fit field strength.\\

\begin{figure*}[]
\begin{center}
\begin{tabular}{ll}

\epsfig{file=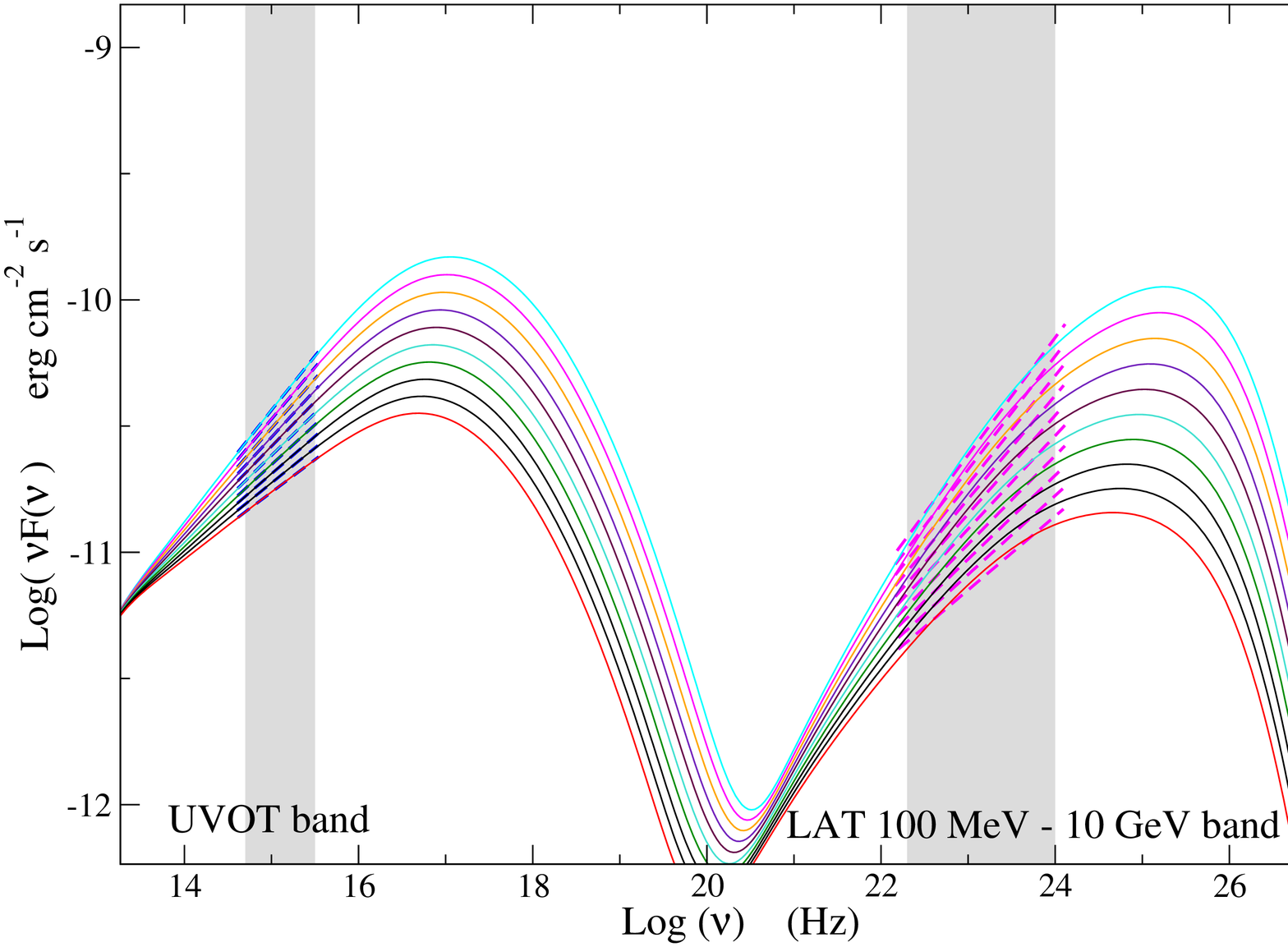 ,width=5cm,angle=-90}
&\epsfig{file=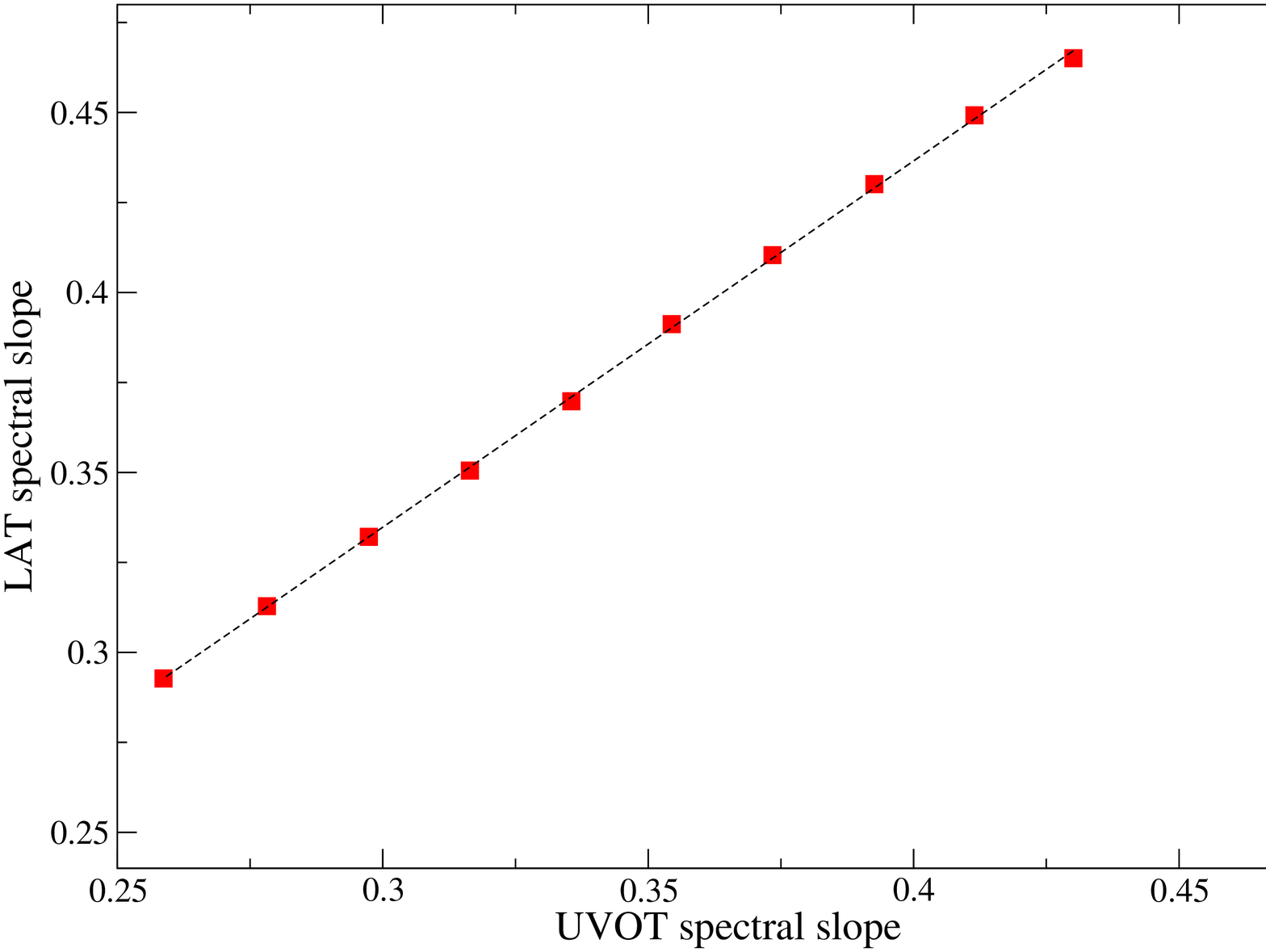 ,width=5cm,angle=-90}\\

\end{tabular}
\end{center}
\caption{The strong correlation between the UVOT spectral slope and that in the LAT band for a SSC
scenario. In the left panel we show the SSC SEDs and in the right panel we show the correlation
between the spectral slope in the \uvt~ band  and that in the \latf~ band. Basic parameters
for the SSC model are: particle number density $N=10 ~cm^{-3}$, magnet	ic field intensity $B=0.1~G$,
minimum electron energy $\gamma_{min}=10^3$, maximum electron energy $\gamma_{max}=5\times 10^6$,
emitting region radius $R=3.0\times15~\mbox{cm}$, beaming factor $\delta=15$, electron curvature $r=1.0$,
power law of index of the low-energy tail of the electron distribution $s=[2.2:2.5]$, and turn-over
energy in the electron distribution $\gamma_c=5.0\times10^4$}
\label{fig-uvot-vs-lat}
\end{figure*}

\subsection{The connection  between the UV slope and that at MeV/GeV energies.}
A further and very relevant consequence of the spectral shape in the UV band is 
its connection with the shape expected at MeV/GeV energies, and in particular in 
the \lat~ band. \\
In order to explore this  we simulated  SSC emission accounting for a 
typical HBL SED. We used as  the electron distribution a power-law at low energy 
with a log-parabolic high energy branch  (LPPL) (Eq. \ref{lppl-elec}). 
The SSC parameter are reported in the caption of Fig. \ref{fig-uvot-vs-lat}. 
In the left panel of this figure we show different SEDs obtained by varying the $s$ 
parameter  in the range [2.1:2.5], and in the right panel we show how  the spectral 
shape in the \uvt~ band is tightly correlated with that in the \latf~ energy range. 
This means that a comparison between the two spectral slopes performed with 
simultaneous observation can hint  whether actually the UV photons from the same 
component emitting in the X-ray are responsible for the SSC emission.

\setcounter{table}{3}
\begin{table*}[htpb]
\caption{SSC best fit results for the 2006 \swf~ observations and using as electron
distribution a log-parabola with a power-law low-energy branch (Eq. \ref{lppl-elec}).}
\label{ssc-fit}
\begin{flushleft}
{\small
\begin{tabular}{llllllllllll}

\hline

\noalign{\smallskip}

Date   &$B$   &$R$   &$\delta$  &$N(*)$       &$r$ &$s$ &$\gamma_{c}(**)$
&$\gamma_{p}^{LP}(***)$  &$\gamma_{max}$
&$\gamma_{min}$ &$u_e{}/u_{b}$\\ 
        &$G$   &$cm$  &          &$cm^{-3}$ &    &    &              &\\ 
\hline
22-04-2006  &$0.1$ &$2.1\times10^{15}$ &25 &13.5   &0.75 &2.30 &$1.75\times10^{5}$
&$2.0\times10^3$ &$2.5\times10^{6}$ &$1.1\times10^{3}$ &119\\

23-06-2006  &$0.1$ &$2.1\times10^{15}$ &25 &15.0   &0.65 &2.30 &$2.85\times10^{5}$
&$4.8\times10^3$ &$4.0\times10^{6}$ &$1.1\times 10^{3}$ &135\\

15-07-2006  &$0.1$ &$2.1\times10^{15}$ &25 &21.0   &0.85 &2.32 &$2.50\times10^{5}$
&$1.0\times10^4$ &$3.0\times10^{6}$ &$1.1\times 10^{3}$ &185\\
 \hline
\noalign{\smallskip} 
\end{tabular}
}
\end{flushleft}
\end{table*}

\setcounter{table}{4}
\begin{table*}[htpb]
\caption{SSC best fit results for  31/03/2005 \swf~ observation and using a  log-parabolic electron
distribution as defined in Eq. \ref{eq-lpep-elec}.}
\label{ssc-fit-1}
\begin{flushleft}
{\small
\begin{tabular}{llllllllll}

\hline

\noalign{\smallskip}

 Date   &$B$   &$R$   &$\delta$  &$N(*)$       &$r$ &$\gamma_{p}(**)$  &$\gamma_{max}$
&$\gamma_{min}$ &$u_e{}/u_{b}$\\ 
        &$G$   &$cm$  &          &$cm^{-3}$ &    &              &\\ 
\hline
31-03-2005  &$0.075$ &$1.5\times10^{15}$ &25 &4   &1.3
&$4.0\times10^{3}$&$2.5\times10^{6}$&$1.1\times10^{3}$& 220\\
\hline
\noalign{\smallskip} 
\end{tabular}
}
\end{flushleft}
(*)  $N=\int n(\gamma)d\gamma$.\\
(**) Do not confuse $\gamma_{c}$ (the turn-over energy in Eq. \ref{lppl-elec})  with 
$\gamma_p$  (the peak energy in Eq. \ref{eq-lpep-elec} ).\\
(***) $\gamma_p^{LP}$ represents $\gamma_p$ obtained from extrapolating the log-parabolic branch of the
distribution described by Eq. \ref{lppl-elec}. These values can be compared with $\gamma_p$ from 31/03/2005.\\
\end{table*}

\section{Discussion.}
In this paper we presented analyses of the spectral and flux 
evolution of \mrk~ during the spring/summer 2006 \swf~ observations, with the 
June pointings monitoring the source almost continuously for 12 days.\\
During this period the source exhibited both flux levels and SED peak energies at 
their historic maximum until 2006.
This very intense flaring state of  \mrk~ represented a unique opportunity to test the 
correlations among the spectral parameters presented in previous works \citep{Trama2007b}, 
expanding the  temporal spanning to about 10 years and enlarging  the volume of the 
parameter  space.\\

The spectral evolution of the source and the patterns in the $a$-flux plane (Sect. 4) 
suggest that each flare has a different characterization  in terms of 
competition among the relevant time scales. Some  flares showed a rise in the hard
and medium X-ray band much faster than that observed in the soft  X-rays. 
This behavior is likely to be explained with the flaring component starting in 
the hard X-ray band and  may suggest that the  driver of those flares is the 
rapid injection of  very energetic particles rather than a gradual acceleration.\\

The $S_p-E_p$ and $E_p-b$ trends follow  those presented in \citet{Trama2007b} with
$E_p$ and $S_p$  correlated and $E_p$ ,$b$ anti-correlated. These trends are
relevant in order to understand the physical mechanisms driving the evolution of 
the electron distribution under the effect of acceleration processes and may 
represent a common scenario for HBLs. In fact, \citet{Massaro2008} confirmed that 
the $E_p-b$ relation holds for five of the TeV HBLs included in their analysis 
(PKS~ 0548-322, 1H~1426+418, Mrk 501, 1ES 1959+650, PKS~2155-304)), and that as for 
the $S_p-E_p$ the only exception to the previous list is PKS~2155-304

The  $E_p-b$ relation shows a possible signature of acceleration processes 
leading to curved electron distributions, with the curvature decreasing as 
the acceleration gets more efficient. 
A first scenario supporting  this picture is that of an  energy dependent 
acceleration probability process \citep{Massaro2004a}.
An alternative explanation is provided by the stochastic acceleration framework, 
with the presence of a momentum-diffusion term \citep{Karda1962,Trama2007b}. 
Both  scenarios predict a negative correlation between $E_p$ and $b$ and 
are consistent with the the data plotted in Fig. \ref{fig-Ep-b}. 
It is worth noting that the data presented here substantially stretch 
the parameter space, confirming that the relation between the curvature and the 
peak energy follows the same trend from the faintest to the strongest flaring 
activity, hinting that the acceleration process is actually be the same.\\
A future investigation may be focused on understanding  which are the physical 
parameters  that can tune the acceleration process. In this regard, an interesting 
investigation may be the connection of the stochastic scenario with the 
turbulence spectrum  of the MHD waves and the resulting diffusion coefficient 
\citep{Park1995, Becker2006, Katar2006, Staw2008}. Understanding the role of the 
turbulence spectrum, and in general, the role of the fluctuation on the particle 
energy gain,  is   a complex task and requires a more detailed analysis that is 
beyond the purpose of this paper and will be presented in Paper II \citep{Tramacere2009}.\\

The $S_p-E_p$ trend demonstrates the connection between the average energy of the particle
distribution and the power output of the source. The expected power-law dependence
$S_p\propto~E_p^{\alpha}$ compared to the observed values of $\alpha$, 
rules out $B,\delta$ and $N$, indicating $\gamma_p$ as the main driver of 
the $S_p-E_p$ trend, confirming the result from \citet{Trama2007b}. A more 
detailed analysis of the scatter plot  reported in Fig.  \ref{fig-Sp-Ep-fit} 
revealed that the trend has a break at about 1 keV, where the typical source luminosity is 
about $L_p\simeq 10^{45}~$ erg/s. This break may be interpreted as the marker of 
the competition between systematic and momentum-diffusion acceleration or in terms 
of energetic content of the jet. We will   study such a scenario in Paper II, 
where we will take  into account also the role $E_p-b$ trend on $S_p-E_p$.\\

Another   interesting result from our analyses is the presence a  power-law tail 
connecting the UV data to the soft-X-ray.
As explained in Sect. 7, we cannot determine whether this tail is a 
characteristic only of highly energetic flares, or whether it is present during 
low states too.  The puzzling aspect in the latter case would be to reconcile
a harder index with a lower flux.
Anyway, if this feature is present only during strong flares then  it may be 
argued that a relevant change has taken place in the acceleration environment.
In fact,  both SA and EDAP scenarios require the presence of a significant 
escape term to develop a power-law tail. \\

The value of the power-law spectral index in the electron distribution $s \simeq 2.3$ 
is very close to the prediction from  relativistic Fermi  first order acceleration 
models \citep{Achterberg2001,Gallant1999,Lemoine2003, Blasi2005,Ellison2004} and 
it is similar to that found in the X-ray observations of GRB afterglows \citep{Waxman1997}. 

This observational feature, which is  in nice agreement with first order acceleration 
models, may suggest a kind of contradiction with the results from the $E_p-b$ trend, 
supporting a stochastic  scenario. A more careful analysis shows that such 
a contradiction is only apparent. In fact, as pointed out recently by \citet{Spit2008}, 
diffusive shock acceleration models rely on the presence of a magnetic turbulence 
near the shock responsible for the scattering process, but these models do not 
clearly  determine the possible role of the turbulence in the acceleration. 
The indication from our observational analysis may be  simultaneous 
roles for the first and second order processes, with the stochastic acceleration 
arising from the magnetic turbulence and signed by the curvature  and more generally 
by  the $E_p-b$ relation, and the first order  signed by the slope of the electron 
power-law tail.\\ 
In such a circumstance the study of the curvature may offer an observational constraint 
for the turbulence properties, and may be recursively used to constraint the scattering 
process  near the shock in the first order models.\\
We note moreover that the model presented by \citet{Spit2008}, in which the shock 
acceleration process is studied in a self-consistent fashion, gives results that 
are compatible with our phenomenological picture. 
The result from this numerical study is a  relativistic 
Maxwellian, and a high-energy tail with $s=2.4\pm0.1$, plus an exponential cut-off 
moving to higher energies with time of the simulation.\\

This phenomenological scenario underlines the relevance of the multi-wavelength 
observations, and warns about the pitfalls of extrapolating the observed spectral 
shape over too wide a range. In this regard we stress the unique capabilities of 
\swf~ to perform simultaneous UV-to-X-ray observations.\\

A further  point concerning the UV observations is given by the expected 
correlation of this spectral band with MeV/GeV band. In fact due to the effect of 
the Klein-Nishina suppression in  the Inverse Compton process, we expect the UV 
photons to be the most efficiently up-scattered at GeV energies by the electrons 
radiating in the X-ray. This makes a strong correlation between the spectral slope in the 
UV-to-soft-X-ray band with the slope at MeV/GeV energies and in particular in the 
\latf~ band. A test of such a correlation would be useful in order to understand 
whether the photons up-scattered at $\gamma-$ray energies are co-spatial with the 
electrons emitting in the X-ray.\\

\section{Conclusions.}
In conclusion the present work shows the complexity of the physical scenario at 
work in the jets of HBL objects. We have shown that the flaring activity of \mrk~  
not only causes  huge flux variation but also results in complex spectral 
evolutions and drastic changes in the electron energy distribution. 
The evolution of the spectral parameters, in particular  $E_p$ and $b$, agrees 
with an acceleration model where the curvature term is inversely proportional 
to $E_p$, which is consistent with both SA and EDAP scenarios. This supports the 
hypothesis that the spectral curvature is related to the acceleration rather 
than to the cooling process. In fact we find less curved spectra as $E_p$ 
increases, leading to the conclusion  that the cooling is more relevant to determine 
the reaching of the equilibrium energy rather than to determine the spectral 
bending.  The presence of a power-law low-energy tail remains puzzling because 
we cannot  determine whether it develops only during strong flares or whether it 
is a common feature. The value $s\simeq 2.3$ is very close to the  \textit{universal} 
index from relativistic shock models and could be used to understand the relative 
weight of the first order process during strong flares, relative to the diffusive 
process invoked to explain the spectral curvature.  We remand these open questions 
to Paper II, were we will focus mainly on the theoretical interpretation of the 
phenomenological picture presented here.\\  As a final remark, we stress that as   
already pointed out in \citet{Massaro2006}, the curvature observed in the X-ray 
spectra of HBLs is reflected in the TeV spectrum. The consequence is that the TeV cut-off 
observed in the nearby HBLs is almost entirely due to the intrinsic electron 
curvature rather than to the interaction with EBL photons. This scenario is 
therefore consistent with a low EBL density that makes the universe more transparent 
to high-energy radiation than previously assumed, in agreement with the discovery 
of HBLs objects at large redshift like  H 2356-309 (z=0.165), 1ES 1101-232 (z=0.186)  
\citep{Aha2006} and 1ES 128-304 (z=0.182) \citep{Albert2006} .

\begin{acknowledgements}
A. Tramacere acknowledges support  by a fellowship of the Italian Space Agency (ASI) and
Istituto Nazionale di Astrofisica (INAF) related to the GLAST Space Mission, through the ASI/INAF ~
I/010/06/0 contract. Gino Tosti  acknowledges support  by the ASI/INAF ~ I/010/06/0 contract.
We thank   Dr. D. Paneque and Dr. S. Digel for  useful comments. 
\end{acknowledgements}

\setcounter{table}{1} 
     \begin{table*}
    \caption{\swx~ Spectral analysis of \mrk.}
     \label{tab-fit} 
     \begin{flushleft} 
     {\small
    \begin{tabular}{llllllllll}
    \hline 
     \hline 
     Orbit &$a$ &$b$ &$K$ &$E_p*$ &$E_p$ &$S_p$                            &flux 0.3-10 keV                &$L_p$                       &$\chi^2_r$/dof \\ 
           &          &         &     &keV    &keV   &\tiny{$10^{-12} erg~ cm^{-2} ~s^{-1}$}  &\tiny{$10^{-12}erg~ cm^{-2} ~s^{-1}$} &\tiny{$10^{45}~erg/s $}     &\\ 
    \hline 
    \noalign{\smallskip}
\hline
\noalign {\smallskip \textit{\textbf{ObsId 00206476000 ~~ Date 4/22/2006 ~~ MJD 53847.252792}}} \\ 
01 &1.69(0.02) &0.09(0.05) &0.347(0.005) &$>10$ &-~-(-~-)  &-~-(-~-)  &2175.7 &-~-(-~-) &0.861(137) \\ 
02 &1.66(0.02) &0.13(0.04) &0.349(0.004) &$>10$ &-~-(-~-)  &-~-(-~-)  &2191.2 &-~-(-~-) &1.181(197) \\ 
03 &1.71(0.02) &0.07(0.04) &0.336(0.004) &$>100$ &-~-(-~-)  &-~-(-~-)  &2119.3 &-~-(-~-) &1.211(196) \\ 
04 &1.67(0.02) &0.12(0.04) &0.334(0.004) &$>10$ &-~-(-~-)  &-~-(-~-)  &2093.5 &-~-(-~-) &1.224(170) \\ 
05 &1.67(0.02) &0.12(0.04) &0.369(0.004) &$>10$ &-~-(-~-)  &-~-(-~-)  &2299.0 &-~-(-~-) &0.902(207) \\ 
06 &1.7(0.02) &0.17(0.04) &0.336(0.004) &7(5) &$7^{+5}_{-2}$  &727(34)  &1980.1 &1.56(0.07) &1.049(192) \\ 
07 &1.69(0.03) &0.21(0.06) &0.306(0.005) &5(4) &$5^{+4}_{-1}$  &635(32)  &1761.1 &1.37(0.07) &1.016(94) \\ 
08 &1.86(0.03) &0.15(0.07) &0.275(0.006) &3(2) &$2.9^{+3}_{-0.8}$  &475(16)  &1452.2 &1.02(0.03) &1.025(58) \\ 
09 &1.84(0.03) &0.14(0.06) &0.25(0.004) &4(3) &$3.6^{+3}_{-0.9}$  &444(14)  &1347.2 &0.96(0.03) &1.116(96) \\ 
10 &1.79(0.03) &0.15(0.06) &0.247(0.004) &5(5) &$5^{+6}_{-2}$  &465(22)  &1367.5 &1(0.05) &0.935(93) \\ 
11 &1.82(0.03) &0.08(0.07) &0.241(0.005) &$>10$ &-~-(-~-)  &-~-(-~-)  &1374.0 &-~-(-~-) &1.163(80) \\ 
12 &1.8(0.03) &0.08(0.06) &0.253(0.004) &$>10$ &-~-(-~-)  &-~-(-~-)  &1447.7 &-~-(-~-) &0.991(98) \\ 
13 &1.78(0.02) &0.09(0.05) &0.269(0.004) &$>10$ &-~-(-~-)  &-~-(-~-)  &1562.1 &-~-(-~-) &1.020(105) \\ 
14 &1.79(0.02) &0.08(0.05) &0.267(0.004) &$>10$ &-~-(-~-)  &-~-(-~-)  &1558.9 &-~-(-~-) &0.989(127) \\ 
15 &1.83(0.02) &0.06(0.05) &0.253(0.004) &$>10$ &-~-(-~-)  &-~-(-~-)  &1445.0 &-~-(-~-) &1.166(119) \\ 
16 &1.81(0.02) &0.07(0.05) &0.269(0.004) &$>10$ &-~-(-~-)  &-~-(-~-)  &1557.7 &-~-(-~-) &1.103(122) \\ 
17 &1.67(0.03) &0.2(0.06) &0.285(0.005) &7(7) &$7^{+8}_{-2}$  &630(47)  &1691.3 &1.4(0.1) &0.892(90) \\ 
18 &1.85(0.03) &0.12(0.06) &0.25(0.005) &5(6) &-~-(-~-)  &-~-(-~-)  &1365.0 &-~-(-~-) &0.863(89) \\ 
19 &1.81(0.03) &0.14(0.06) &0.246(0.005) &5(6) &-~-(-~-)  &-~-(-~-)  &1362.0 &-~-(-~-) &1.218(85) \\ 
20 &1.83(0.02) &0.12(0.05) &0.251(0.004) &5(6) &-~-(-~-)  &-~-(-~-)  &1382.7 &-~-(-~-) &0.945(115) \\ 
21 &1.81(0.02) &0.17(0.05) &0.276(0.004) &4(2) &$3.8^{+2}_{-0.8}$  &502(14)  &1500.4 &1.08(0.03) &1.083(117) \\ 
22 &1.83(0.02) &0.09(0.05) &0.294(0.004) &9(15) &-~-(-~-)  &-~-(-~-)  &1649.9 &-~-(-~-) &1.127(130) \\ 
23 &1.77(0.02) &0.12(0.05) &0.275(0.004) &9(13) &-~-(-~-)  &-~-(-~-)  &1586.6 &-~-(-~-) &1.044(122) \\ 
24 &1.82(0.02) &0.09(0.05) &0.267(0.004) &9(15) &-~-(-~-)  &-~-(-~-)  &1505.8 &-~-(-~-) &1.066(113) \\ 
25 &1.8(0.02) &0.12(0.05) &0.283(0.004) &7(8) &$7^{+12}_{-2}$  &548(34)  &1589.9 &1.18(0.07) &0.901(132) \\ 
26 &1.8(0.02) &0.12(0.05) &0.282(0.004) &6(7) &-~-(-~-)  &-~-(-~-)  &1579.1 &-~-(-~-) &0.994(125) \\ 
27 &1.74(0.03) &0.14(0.07) &0.283(0.006) &8(14) &-~-(-~-)  &-~-(-~-)  &1636.3 &-~-(-~-) &1.120(64) \\ 
\hline
\noalign {\smallskip \textit{\textbf{ObsId 00030352005 ~~ Date 4/25/2006 ~~ MJD 53850.267940}}} \\ 
01 &1.84(0.01) &0.15(0.02) &0.273(0.002) &3(1) &$3.3^{+0.7}_{-0.4}$  &481(5)  &1468.4 &1.03(0.01) &1.360(288) \\ 
02 &1.86(0.01) &0.16(0.02) &0.264(0.002) &2.7(0.7) &$2.7^{+0.4}_{-0.3}$  &454(4)  &1391.7 &0.977(0.009) &1.248(285) \\ 
03 &1.86(0.01) &0.15(0.02) &0.269(0.002) &2.9(0.8) &$2.9^{+0.5}_{-0.3}$  &464(4)  &1427.2 &0.998(0.009) &1.090(287) \\ 
\hline
\noalign {\smallskip \textit{\textbf{ObsId 00030352006 ~~ Date 4/26/2006 ~~ MJD 53851.146539}}} \\ 
01(*) &1.86(0.02) &0.17(0.03) &0.42(0.004) &2.5(0.7) &$2.5^{+0.5}_{-0.3}$  &715(9)  &2192.5
&1.54(0.02) &0.989(208) \\ 
02 &1.83(0.01) &0.14(0.03) &0.285(0.002) &4(2) &$4.2^{+2}_{-0.8}$  &515(9)  &1556.8 &1.11(0.02) &1.257(253) \\ 
03 &1.82(0.01) &0.17(0.03) &0.294(0.002) &3(1) &$3.4^{+0.7}_{-0.4}$  &525(7)  &1584.4 &1.13(0.01) &1.380(254) \\ 
\hline
\noalign {\smallskip \textit{\textbf{ObsId 00030352007 ~~ Date 4/26/2006 ~~ MJD 53851.951963}}} \\ 
01 &1.87(0.01) &0.16(0.03) &0.265(0.002) &2.6(0.6) &$2.6^{+0.4}_{-0.3}$  &451(4)  &1388.5 &0.97(0.009) &1.140(258) \\ 
\hline
\noalign {\smallskip \textit{\textbf{ObsId 00030352008 ~~ Date 6/14/2006 ~~ MJD 53900.016100}}} \\ 
01 &2(0.03) &0.12(0.07) &0.25(0.005) &1(0.2) &$1^{+0.2}_{-0.4}$  &400(8)  &1227.0 &0.86(0.02) &0.924(178) \\ 
02 &2.07(0.03) &0.14(0.06) &0.236(0.004) &0.6(0.2) &$0.6^{+0.2}_{-0.3}$  &384(10)  &1106.2 &0.83(0.02) &1.036(173) \\ 
03 &2.07(0.03) &0.05(0.07) &0.23(0.005) &0.2(0.7) &-~-(-~-)  &-~-(-~-)  &1130.2 &-~-(-~-) &0.862(171) \\ 
04 &2.01(0.03) &0.2(0.07) &0.236(0.005) &0.9(0.1) &$0.9^{+0.1}_{-0.2}$  &378(7)  &1111.7 &0.81(0.01) &1.006(175) \\ 
05 &2.1(0.03) &0.2(0.06) &0.213(0.004) &0.6(0.2) &$0.6^{+0.1}_{-0.2}$  &351(8)  &960.3 &0.76(0.02) &1.089(175) \\ 
\hline
    \noalign{\smallskip}
    \end{tabular}
    }
    \end{flushleft}
(*) gti with biased exposure, the affected data are not taken into account in the analyses.\\
The second third and forth columns report the best fit estimate for the model
in Eq. \ref{eq-lp}. The fifth column reports the value of the SED peak analytically  estimated
from Eq. \ref{eq-lp} according to the best fit results. The sixth and seventh columns report the
$E_p$ and $S_p$ best fit estimates using as best fit model Eq. \ref{eq-lpep}. In the eighth column
we report the flux in the 0.3-10.0 keV band, evaluated by Xspec integrating the Eq. \ref{eq-lp} model.
The ninth colum reports the SED peak flux luminosity evaluated as 
$L_p~\simeq~S_p~4\pi~ D_L^2$, where $D_L\simeq 134.1$ Mpc is the luminosity distance.
In the last column we report the reduced $\chi^2$ and the degrees of freedom concerning the  Eq. \ref{eq-lp}  
fit.
    \end{table*}

\setcounter{table}{1} 
     \begin{table*} 
     \caption{\swx~ Spectral analysis of \mrk. \it{continued}} 
     \label{tab-fit} 
     \begin{flushleft} 
     {\small
    \begin{tabular}{llllllllll}
    \hline 
    \hline 
    Orbit  &$a$ &$b$ &$K$ &$E_p*$ &$E_p$ &$S_p$                            &flux 0.3-10 keV                 &$L_p$                     &$\chi^2_r$/dof \\ 
           &          &         &     &keV    &keV   &\tiny{$10^{-12} erg~ cm^{-2} ~s^{-1}$}  &\tiny{$10^{-12}erg~ cm^{-2} ~s^{-1}$} &\tiny{$10^{45}~erg/s$}    &\\ 
    \hline 
    \noalign{\smallskip}
06 &2.1(0.03) &0.06(0.07) &0.195(0.004) &0.2(0.5) &$0.16^{+0.1}_{-0.07}$  &343(18)  &939.1 &0.74(0.04) &1.099(147) \\ 
07 &2.08(0.03) &0.07(0.07) &0.197(0.004) &0.3(0.6) &$0.27^{+0.2}_{-0.07}$  &332(14)  &960.0 &0.71(0.03) &0.873(143) \\ 
08 &2.05(0.04) &-0.08(0.09) &0.192(0.005) &-~- &-~-(-~-)  &-~-(-~-)  &1028.5 &-~-(-~-) &1.038(96) \\ 
09 &2.02(0.02) &-0.01(0.04) &0.167(0.002) &-~- &-~-(-~-)  &-~-(-~-)  &872.3 &-~-(-~-) &1.137(266) \\ 
\hline
\noalign {\smallskip \textit{\textbf{ObsId 00030352009 ~~ Date 6/15/2006 ~~ MJD 53901.490493}}} \\ 
01 &1.83(0.02) &0.06(0.03) &0.181(0.002) &$>10$ &-~-(-~-)  &-~-(-~-)  &1035.8 &-~-(-~-) &0.979(270) \\ 
02 &1.89(0.02) &-0.02(0.04) &0.162(0.002) &-~- &-~-(-~-)  &-~-(-~-)  &941.5 &-~-(-~-) &1.150(219) \\ 
03 &1.87(0.03) &0.01(0.06) &0.14(0.002) &$>100$ &-~-(-~-)  &-~-(-~-)  &803.3 &-~-(-~-) &0.842(155) \\ 
04 &1.86(0.02) &0.04(0.05) &0.194(0.003) &$>10$ &-~-(-~-)  &-~-(-~-)  &1100.3 &-~-(-~-) &0.936(200) \\ 
05 &1.82(0.03) &0.13(0.07) &0.196(0.004) &5(6) &-~-(-~-)  &-~-(-~-)  &1078.0 &-~-(-~-) &0.930(111) \\ 
06 &1.88(0.03) &0.11(0.07) &0.191(0.004) &4(5) &-~-(-~-)  &-~-(-~-)  &1021.2 &-~-(-~-) &1.313(97) \\ 
07 &1.88(0.01) &0.16(0.03) &0.19(0.002) &2.3(0.7) &$2.3^{+0.4}_{-0.3}$  &319(4)  &985.1 &0.686(0.009) &1.017(273) \\ 
\hline
\noalign {\smallskip \textit{\textbf{ObsId 00030352010 ~~ Date 6/16/2006 ~~ MJD 53902.025296}}} \\ 
01 &1.92(0.01) &0.08(0.03) &0.201(0.002) &3(2) &$3.2^{+2}_{-0.8}$  &337(5)  &1066.9 &0.73(0.01) &1.287(222) \\ 
02 &1.84(0.01) &0.1(0.03) &0.219(0.002) &6(5) &$6^{+6}_{-2}$  &404(13)  &1213.1 &0.87(0.03) &1.228(237) \\ 
03 &1.85(0.01) &0.1(0.03) &0.233(0.002) &6(4) &$6^{+5}_{-2}$  &425(12)  &1283.5 &0.91(0.02) &1.142(242) \\ 
04 &1.83(0.01) &0.07(0.03) &0.254(0.003) &$>10$ &-~-(-~-)  &-~-(-~-)  &1446.7 &-~-(-~-) &1.234(226) \\ 
05 &1.79(0.01) &0.1(0.03) &0.263(0.002) &$>10$ &$12^{+17}_{-5}$  &546(32)  &1516.0 &1.17(0.07) &1.146(263) \\ 
06 &1.77(0.01) &0.1(0.03) &0.268(0.002) &$>10$ &$16^{+25}_{-7}$  &590(43)  &1574.0 &1.27(0.09) &1.227(263) \\ 
07 &1.73(0.02) &0.02(0.04) &0.279(0.003) &$>100$ &-~-(-~-)  &-~-(-~-)  &1773.0 &-~-(-~-) &1.256(210) \\ 
08 &1.76(0.02) &-0.03(0.05) &0.281(0.004) &-~- &-~-(-~-)  &-~-(-~-)  &1808.0 &-~-(-~-) &0.926(111) \\ 
09(*) &1.72(0.02) &0.02(0.03) &0.427(0.004) &$>100$ &-~-(-~-)  &-~-(-~-)  &2741.8 &-~-(-~-)
&1.180(230) \\ 
10 &1.69(0.02) &0.04(0.03) &0.261(0.003) &$>100$ &-~-(-~-)  &-~-(-~-)  &1697.5 &-~-(-~-) &1.110(220) \\ 
11 &1.7(0.02) &0.04(0.03) &0.285(0.003) &$>100$ &-~-(-~-)  &-~-(-~-)  &1833.8 &-~-(-~-) &1.139(227) \\ 
12 &1.73(0.01) &0.03(0.03) &0.302(0.003) &$>100$ &-~-(-~-)  &-~-(-~-)  &1911.1 &-~-(-~-) &1.258(240) \\ 
13 &1.76(0.02) &0(0.03) &0.279(0.003) &-~- &-~-(-~-)  &-~-(-~-)  &1764.0 &-~-(-~-) &1.120(230) \\ 
14 &1.74(0.01) &0.04(0.02) &0.29(0.002) &$>100$ &-~-(-~-)  &-~-(-~-)  &1812.6 &-~-(-~-) &1.034(288) \\ 
15 &1.77(0.01) &0.04(0.02) &0.277(0.002) &$>100$ &-~-(-~-)  &-~-(-~-)  &1677.1 &-~-(-~-) &1.093(279) \\ 
16 &1.81(0.01) &0.05(0.03) &0.246(0.002) &$>10$ &-~-(-~-)  &-~-(-~-)  &1443.2 &-~-(-~-) &1.128(242) \\ 
17 &1.8(0.01) &0.03(0.03) &0.248(0.002) &$>100$ &-~-(-~-)  &-~-(-~-)  &1485.6 &-~-(-~-) &1.330(262) \\ 
18 &1.84(0.01) &-0.05(0.03) &0.248(0.002) &-~- &-~-(-~-)  &-~-(-~-)  &1512.6 &-~-(-~-) &1.049(236) \\ 
19 &1.84(0.01) &-0.05(0.03) &0.225(0.002) &-~- &-~-(-~-)  &-~-(-~-)  &1383.1 &-~-(-~-) &0.978(221) \\ 
20 &1.81(0.01) &-0.04(0.03) &0.245(0.002) &-~- &-~-(-~-)  &-~-(-~-)  &1517.3 &-~-(-~-) &1.100(241) \\ 
21 &1.7(0.02) &0.03(0.03) &0.287(0.003) &$>100$ &-~-(-~-)  &-~-(-~-)  &1871.2 &-~-(-~-) &1.038(240) \\ 
22 &1.77(0.01) &0(0.03) &0.284(0.003) &-~- &-~-(-~-)  &-~-(-~-)  &1769.6 &-~-(-~-) &1.214(234) \\ 
23 &1.75(0.02) &0.03(0.03) &0.293(0.003) &$>100$ &-~-(-~-)  &-~-(-~-)  &1833.7 &-~-(-~-) &1.116(226) \\ 
24 &1.77(0.02) &0.05(0.03) &0.32(0.003) &$>100$ &-~-(-~-)  &-~-(-~-)  &1930.8 &-~-(-~-) &1.301(215) \\ 
\hline
\noalign {\smallskip \textit{\textbf{ObsId 00030352011 ~~ Date 6/18/2006 ~~ MJD 53904.038766}}} \\ 
01 &1.8(0.01) &0.11(0.03) &0.318(0.003) &9(8) &$9^{+8}_{-3}$  &633(27)  &1808.8 &1.36(0.06) &1.036(286) \\ 
02 &1.82(0.01) &0.12(0.03) &0.319(0.003) &6(4) &$6^{+3}_{-1}$  &600(15)  &1777.2 &1.29(0.03) &1.227(286) \\ 
03 &1.82(0.01) &0.18(0.02) &0.34(0.002) &3.1(0.7) &$3.1^{+0.4}_{-0.3}$  &601(6)  &1817.5 &1.29(0.01) &1.343(317) \\ 
04 &1.93(0.01) &0.13(0.02) &0.294(0.002) &1.9(0.4) &$1.9^{+0.2}_{-0.2}$  &482(4)  &1511.3 &1.037(0.009) &1.211(297) \\ 
05 &1.94(0.01) &0.11(0.03) &0.292(0.002) &1.9(0.5) &$1.9^{+0.3}_{-0.2}$  &476(4)  &1507.0 &1.024(0.009) &1.216(287) \\ 
06 &2.03(0.02) &0.04(0.04) &0.292(0.004) &0.5(0.5) &$0.5^{+0.4}_{-0.5}$  &471(24)  &1482.8 &1.01(0.05) &1.242(198) \\ 
\hline
    \noalign{\smallskip}
    \end{tabular}
    }
    \end{flushleft}
(*) gti with biased exposure, the affected data are not taken into account in the analyses.\\
The second third and forth columns report the best fit estimate for the model
in Eq. \ref{eq-lp}. The fifth column reports the value of the SED peak analytically  estimated
from Eq. \ref{eq-lp} according to the best fit results. The sixth and seventh columns report the
$E_p$ and $S_p$ best fit estimates using as best fit model Eq. \ref{eq-lpep}. In the eighth column
we report the flux in the 0.3-10.0 keV band, evaluated by Xspec integrating the Eq. \ref{eq-lp} model.
The ninth colum reports the SED peak flux luminosity evaluated as 
$L_p~\simeq~S_p~4\pi~ D_L^2$, where $D_L\simeq 134.1$ Mpc is the luminosity distance.
In the last column we report the reduced $\chi^2$ and the degrees of freedom concerning the  Eq. \ref{eq-lp}  
fit.
    \end{table*}

\setcounter{table}{1} 
     \begin{table*} 
     \caption{\swx~ Spectral analysis of \mrk. \it{continued}} 
     \label{tab-fit} 
     \begin{flushleft} 
     {\small
    \begin{tabular}{llllllllll}
    \hline 
    \hline 
    Orbit  &$a$ &$b$ &K &$E_p*$ &$E_p$ &$S_p$                            &flux 0.3-10 keV                 &$L_p$                     &$\chi^2_r$/dof \\ 
           &          &         &     &keV    &keV   &\tiny{$10^{-12} erg~ cm^{-2} ~s^{-1}$}  &\tiny{$10^{-12}erg~ cm^{-2} ~s^{-1}$} &\tiny{$10^{45}~erg/s$}    &\\ 
    \hline 
    \noalign{\smallskip}
07 &1.98(0.02) &0.08(0.04) &0.259(0.003) &1.5(0.5) &$1.4^{+0.4}_{-0.3}$  &416(5)  &1330.3 &0.9(0.01) &1.002(213) \\ 
08 &1.96(0.02) &0.15(0.04) &0.267(0.003) &1.4(0.2) &$1.4^{+0.1}_{-0.1}$  &430(5)  &1333.3 &0.93(0.01) &1.199(219) \\ 
09 &1.92(0.02) &0.09(0.03) &0.269(0.003) &3(2) &$2.7^{+2}_{-0.6}$  &448(6)  &1420.2 &0.96(0.01) &1.085(230) \\ 
10 &1.89(0.01) &0.06(0.03) &0.273(0.002) &9(14) &-~-(-~-)  &-~-(-~-)  &1501.8 &-~-(-~-) &1.174(291) \\ 
11 &1.87(0.02) &0.07(0.03) &0.289(0.003) &10(19) &-~-(-~-)  &-~-(-~-)  &1600.5 &-~-(-~-) &1.178(233) \\ 
12 &1.84(0.02) &0.01(0.03) &0.298(0.003) &$>100$ &-~-(-~-)  &-~-(-~-)  &1746.9 &-~-(-~-) &1.016(243) \\ 
13 &1.84(0.02) &0.07(0.03) &0.299(0.003) &$>10$ &-~-(-~-)  &-~-(-~-)  &1687.1 &-~-(-~-) &1.122(240) \\ 
14 &1.85(0.02) &0.09(0.03) &0.286(0.003) &6(7) &$6^{+10}_{-2}$  &525(22)  &1580.9 &1.13(0.05) &1.058(233) \\ 
15 &1.93(0.02) &0.11(0.03) &0.256(0.003) &2(0.7) &$2^{+0.5}_{-0.3}$  &419(5)  &1321.3 &0.9(0.01) &1.228(222) \\ 
16 &1.95(0.02) &0.1(0.04) &0.248(0.003) &1.9(0.8) &$1.9^{+0.6}_{-0.3}$  &403(5)  &1280.6 &0.87(0.01) &1.101(210) \\ 
17 &1.92(0.02) &0.14(0.04) &0.236(0.003) &2(0.6) &$1.9^{+0.4}_{-0.2}$  &388(5)  &1249.5 &0.83(0.01) &1.274(207) \\ 
18 &1.9(0.01) &0.08(0.03) &0.255(0.002) &5(5) &$5^{+6}_{-1}$  &441(11)  &1376.8 &0.95(0.02) &1.112(254) \\ 
19 &1.88(0.01) &0.07(0.03) &0.256(0.002) &7(9) &-~-(-~-)  &-~-(-~-)  &1409.5 &-~-(-~-) &1.088(251) \\ 
20 &1.9(0.01) &0.1(0.03) &0.239(0.002) &3(2) &$3.2^{+2}_{-0.7}$  &407(6)  &1272.2 &0.88(0.01) &1.213(244) \\ 
21 &1.97(0.01) &0.07(0.03) &0.218(0.002) &1.5(0.5) &$1.5^{+0.4}_{-0.3}$  &350(4)  &1121.5 &0.753(0.009) &1.109(230) \\ 
22 &1.99(0.02) &0.11(0.03) &0.204(0.002) &1.2(0.2) &$1.2^{+0.2}_{-0.2}$  &328(4)  &1023.7 &0.706(0.009) &0.993(225) \\ 
23 &1.98(0.01) &0.06(0.03) &0.212(0.002) &1.4(0.6) &$1.4^{+0.6}_{-0.3}$  &340(4)  &1097.1 &0.732(0.009) &0.944(227) \\ 
24 &2.01(0.01) &0.09(0.04) &0.215(0.002) &0.8(0.2) &$0.9^{+0.2}_{-0.2}$  &344(4)  &1072.8 &0.74(0.009) &1.065(220) \\ 
25 &1.98(0.01) &0.03(0.03) &0.234(0.002) &2(2) &-~-(-~-)  &-~-(-~-)  &1229.9 &-~-(-~-) &1.127(235) \\ 
26 &1.85(0.01) &-0.02(0.03) &0.295(0.003) &-~- &-~-(-~-)  &-~-(-~-)  &1809.2 &-~-(-~-) &1.290(269) \\ 
27 &1.83(0.01) &0.02(0.03) &0.285(0.003) &$>100$ &-~-(-~-)  &-~-(-~-)  &1676.1 &-~-(-~-) &1.249(268) \\ 
28 &1.85(0.01) &0.02(0.03) &0.276(0.003) &$>100$ &-~-(-~-)  &-~-(-~-)  &1595.7 &-~-(-~-) &1.322(261) \\ 
29 &1.8(0.01) &0.06(0.03) &0.285(0.002) &$>10$ &-~-(-~-)  &-~-(-~-)  &1662.8 &-~-(-~-) &1.132(293) \\ 
30 &1.83(0.01) &0.02(0.03) &0.29(0.002) &$>100$ &-~-(-~-)  &-~-(-~-)  &1701.8 &-~-(-~-) &1.178(284) \\ 
31 &1.83(0.02) &0.08(0.03) &0.401(0.004) &10(15) &-~-(-~-)  &-~-(-~-)  &2260.2 &-~-(-~-) &1.022(237) \\ 
32 &1.83(0.02) &0.08(0.03) &0.281(0.003) &$>10$ &-~-(-~-)  &-~-(-~-)  &1598.0 &-~-(-~-) &0.894(231) \\ 
33 &1.84(0.02) &0.11(0.04) &0.289(0.003) &6(5) &$6^{+6}_{-2}$  &531(19)  &1593.0 &1.14(0.04) &0.876(220) \\ 
34 &1.82(0.03) &0.17(0.06) &0.29(0.005) &3(2) &$3.5^{+3}_{-0.8}$  &520(16)  &1557.4 &1.12(0.03) &1.092(101) \\ 
\hline
\noalign {\smallskip \textit{\textbf{ObsId 00030352012 ~~ Date 6/20/2006 ~~ MJD 53907.001774}}} \\ 
01 &1.92(0.03) &0.09(0.08) &0.311(0.007) &3(4) &-~-(-~-)  &-~-(-~-)  &1629.2 &-~-(-~-) &0.853(64) \\ 
02 &1.83(0.01) &0.16(0.02) &0.303(0.002) &3.4(0.9) &$3.4^{+0.6}_{-0.4}$  &538(6)  &1631.2 &1.16(0.01) &1.179(297) \\ 
03 &1.83(0.01) &0.16(0.03) &0.286(0.002) &3(1) &$3.3^{+0.7}_{-0.4}$  &506(7)  &1535.8 &1.09(0.01) &1.065(247) \\ 
04 &1.87(0.01) &0.16(0.03) &0.267(0.002) &2.6(0.8) &$2.6^{+0.5}_{-0.3}$  &455(5)  &1399.4 &0.98(0.01) &1.113(235) \\ 
05 &1.87(0.01) &0.18(0.03) &0.253(0.002) &2.3(0.6) &$2.3^{+0.3}_{-0.2}$  &429(5)  &1313.9 &0.92(0.01) &1.163(227) \\ 
06 &1.83(0.02) &-0.01(0.03) &0.198(0.002) &-~- &-~-(-~-)  &-~-(-~-)  &1191.2 &-~-(-~-) &0.932(218) \\ 
07 &1.92(0.02) &0.09(0.04) &0.225(0.003) &3(2) &$2.8^{+2}_{-0.7}$  &376(6)  &1189.5 &0.81(0.01) &0.968(187) \\ 
08 &1.9(0.02) &0.14(0.05) &0.244(0.003) &2(1) &$2.3^{+0.9}_{-0.4}$  &408(7)  &1269.1 &0.88(0.01) &1.046(149) \\ 
09 &1.91(0.02) &0.16(0.05) &0.255(0.004) &2(0.8) &$2^{+0.6}_{-0.3}$  &422(7)  &1305.3 &0.91(0.01) &1.071(112) \\ 
10 &1.88(0.01) &0.15(0.03) &0.275(0.002) &2.5(0.8) &$2.5^{+0.5}_{-0.3}$  &464(5)  &1438.3 &1(0.01) &1.145(239) \\ 
11 &1.85(0.01) &0.14(0.03) &0.294(0.003) &3(2) &$3.4^{+1}_{-0.6}$  &514(8)  &1577.8 &1.11(0.02) &1.284(229) \\ 
12 &1.83(0.01) &0.2(0.03) &0.316(0.003) &2.6(0.6) &$2.6^{+0.4}_{-0.2}$  &548(6)  &1657.9 &1.18(0.01) &1.255(232) \\ 
13 &1.87(0.01) &0.15(0.03) &0.325(0.003) &2.8(0.8) &$2.8^{+0.5}_{-0.3}$  &556(6)  &1713.4 &1.2(0.01) &1.082(262) \\ 
14 &1.85(0.01) &0.19(0.03) &0.337(0.003) &2.5(0.6) &$2.5^{+0.3}_{-0.2}$  &578(6)  &1762.0 &1.24(0.01) &1.167(258) \\ 
15 &1.91(0.01) &0.16(0.03) &0.319(0.003) &1.8(0.3) &$1.8^{+0.2}_{-0.1}$  &524(5)  &1620.8 &1.13(0.01) &1.199(252) \\ 
\hline
    \noalign{\smallskip}
    \end{tabular}
    }
    \end{flushleft}
(*) gti with biased exposure, the affected data are not taken into account in the analyses.\\
The second third and forth columns report the best fit estimate for the model
in Eq. \ref{eq-lp}. The fifth column reports the value of the SED peak analytically  estimated
from Eq. \ref{eq-lp} according to the best fit results. The sixth and seventh columns report the
$E_p$ and $S_p$ best fit estimates using as best fit model Eq. \ref{eq-lpep}. In the eighth column
we report the flux in the 0.3-10.0 keV band, evaluated by Xspec integrating the Eq. \ref{eq-lp} model.
The ninth colum reports the SED peak flux luminosity evaluated as 
$L_p~\simeq~S_p~4\pi~ D_L^2$, where $D_L\simeq 134.1$ Mpc is the luminosity distance.
In the last column we report the reduced $\chi^2$ and the degrees of freedom concerning the  Eq. \ref{eq-lp}  
fit.
    \end{table*}

\setcounter{table}{1} 
     \begin{table*} 
     \caption{\swx~ Spectral analysis of \mrk. \it{continued}} 
     \label{tab-fit} 
     \begin{flushleft} 
     {\small
    \begin{tabular}{llllllllll}
    \hline 
    \hline 
    Orbit  &$a$ &$b$ &K &$E_p*$ &$E_p$ &$S_p$                            &flux 0.3-10 keV                 &$L_p$                     &$\chi^2_r$/dof \\ 
           &          &         &     &keV    &keV   &\tiny{$10^{-12} erg~ cm^{-2} ~s^{-1}$}  &\tiny{$10^{-12}erg~ cm^{-2} ~s^{-1}$} &\tiny{$10^{45}~erg/s$}    &\\ 
    \hline 
    \noalign{\smallskip}
16 &1.92(0.01) &0.14(0.03) &0.316(0.003) &2(0.5) &$2^{+0.3}_{-0.2}$  &519(5)  &1623.1 &1.12(0.01) &1.068(246) \\ 
\hline
\noalign {\smallskip \textit{\textbf{ObsId 00030352013 ~~ Date 6/22/2006 ~~ MJD 53908.050644}}} \\ 
01 &1.96(0.01) &0.18(0.03) &0.291(0.002) &1.3(0.1) &$1.28^{+0.08}_{-0.08}$  &468(4)  &1424.1 &1.007(0.009) &1.191(245) \\ 
02 &1.93(0.01) &0.15(0.03) &0.302(0.002) &1.8(0.3) &$1.8^{+0.2}_{-0.1}$  &493(4)  &1537.5 &1.061(0.009) &1.360(256) \\ 
03 &1.94(0.01) &0.18(0.03) &0.308(0.002) &1.5(0.2) &$1.45^{+0.1}_{-0.09}$  &498(4)  &1522.5 &1.072(0.009) &1.088(253) \\ 
04 &1.98(0.01) &0.15(0.03) &0.269(0.002) &1.2(0.1) &$1.2^{+0.1}_{-0.1}$  &431(4)  &1323.0 &0.927(0.009) &1.083(241) \\ 
05 &2.01(0.01) &0.08(0.03) &0.252(0.002) &0.9(0.2) &$0.9^{+0.2}_{-0.3}$  &404(4)  &1268.0 &0.869(0.009) &1.006(213) \\ 
06 &1.96(0.01) &0.12(0.03) &0.243(0.002) &1.5(0.3) &$1.5^{+0.2}_{-0.1}$  &392(4)  &1231.3 &0.843(0.009) &1.208(235) \\ 
07 &1.93(0.01) &0.03(0.03) &0.243(0.002) &$>10$ &-~-(-~-)  &-~-(-~-)  &1322.9 &-~-(-~-) &1.063(245) \\ 
08 &1.82(0.01) &0.03(0.02) &0.281(0.002) &$>100$ &-~-(-~-)  &-~-(-~-)  &1667.5 &-~-(-~-) &1.434(277) \\ 
09 &1.79(0.01) &0.09(0.03) &0.263(0.002) &$>10$ &$12^{+16}_{-5}$  &545(32)  &1516.0 &1.17(0.07) &1.163(265) \\ 
10 &1.82(0.01) &0.07(0.03) &0.241(0.002) &$>10$ &-~-(-~-)  &-~-(-~-)  &1384.9 &-~-(-~-) &1.275(256) \\ 
11 &1.88(0.01) &0.02(0.03) &0.241(0.002) &$>100$ &-~-(-~-)  &-~-(-~-)  &1374.1 &-~-(-~-) &1.247(256) \\ 
12 &1.88(0.01) &0.02(0.03) &0.229(0.002) &$>100$ &-~-(-~-)  &-~-(-~-)  &1300.0 &-~-(-~-) &1.051(244) \\ 
13 &1.86(0.01) &0.02(0.03) &0.25(0.002) &$>100$ &-~-(-~-)  &-~-(-~-)  &1438.9 &-~-(-~-) &1.036(255) \\ 
14 &1.88(0.01) &0.02(0.03) &0.245(0.002) &$>100$ &-~-(-~-)  &-~-(-~-)  &1393.2 &-~-(-~-) &1.221(254) \\ 
15 &1.88(0.01) &0.12(0.02) &0.231(0.001) &3(1) &$3.1^{+0.7}_{-0.4}$  &395(4)  &1227.7 &0.85(0.009) &1.334(308) \\ 
\hline
\noalign {\smallskip \textit{\textbf{ObsId 00030352014 ~~ Date 6/23/2006 ~~ MJD 53909.394900}}} \\ 
01 &1.74(0.01) &0.02(0.02) &0.326(0.003) &$>100$ &-~-(-~-)  &-~-(-~-)  &2060.1 &-~-(-~-) &1.163(281) \\ 
02 &1.67(0.01) &0.06(0.03) &0.372(0.003) &$>100$ &-~-(-~-)  &-~-(-~-)  &2435.1 &-~-(-~-) &1.240(255) \\ 
03 &1.66(0.01) &0.02(0.03) &0.411(0.004) &$>100$ &-~-(-~-)  &-~-(-~-)  &2783.6 &-~-(-~-) &1.110(276) \\ 
04 &1.69(0.01) &0.04(0.03) &0.398(0.003) &$>100$ &-~-(-~-)  &-~-(-~-)  &2603.6 &-~-(-~-) &1.234(265) \\ 
05 &1.63(0.01) &0.08(0.03) &0.436(0.004) &$>100$ &-~-(-~-)  &-~-(-~-)  &2938.2 &-~-(-~-) &1.357(284) \\ 
06 &1.67(0.01) &0.06(0.02) &0.434(0.003) &$>100$ &-~-(-~-)  &-~-(-~-)  &2845.1 &-~-(-~-) &1.079(291) \\ 
07 &1.68(0.01) &0.07(0.03) &0.444(0.004) &$>100$ &-~-(-~-)  &-~-(-~-)  &2870.2 &-~-(-~-) &1.045(282) \\ 
08 &1.76(0.01) &0.07(0.03) &0.393(0.003) &$>10$ &-~-(-~-)  &-~-(-~-)  &2366.8 &-~-(-~-) &1.098(255) \\ 
09 &1.8(0.01) &0.03(0.02) &0.383(0.003) &$>100$ &-~-(-~-)  &-~-(-~-)  &2302.1 &-~-(-~-) &1.041(286) \\ 
10 &1.79(0.01) &0.06(0.03) &0.384(0.003) &$>10$ &-~-(-~-)  &-~-(-~-)  &2259.2 &-~-(-~-) &1.025(249) \\ 
\hline
\noalign {\smallskip \textit{\textbf{ObsId 00215769000 ~~ Date 6/23/2006 ~~ MJD 53909.671942}}} \\ 
01 &1.64(0.01) &0.05(0.02) &0.386(0.003) &$>100$ &-~-(-~-)  &-~-(-~-)  &2617.0 &-~-(-~-) &1.099(310) \\ 
\hline
\noalign {\smallskip \textit{\textbf{ObsId 00030352015 ~~ Date 6/24/2006 ~~ MJD 53910.068887}}} \\ 
01 &1.83(0.01) &0.04(0.03) &0.447(0.004) &$>100$ &-~-(-~-)  &-~-(-~-)  &2601.0 &-~-(-~-) &1.103(252) \\ 
02 &1.85(0.01) &0.02(0.03) &0.409(0.003) &$>100$ &-~-(-~-)  &-~-(-~-)  &2367.8 &-~-(-~-) &1.206(263) \\ 
03 &1.85(0.01) &0.04(0.03) &0.406(0.003) &$>10$ &-~-(-~-)  &-~-(-~-)  &2323.1 &-~-(-~-) &1.157(263) \\ 
04 &1.83(0.01) &0.07(0.03) &0.418(0.003) &$>10$ &-~-(-~-)  &-~-(-~-)  &2389.7 &-~-(-~-) &1.229(259) \\ 
05 &1.84(0.01) &0.1(0.02) &0.439(0.003) &6(4) &$6^{+4}_{-2}$  &816(22)  &2440.4 &1.76(0.05) &1.071(271) \\ 
06 &1.78(0.01) &0.12(0.03) &0.473(0.004) &9(7) &$9^{+7}_{-3}$  &957(40)  &2707.7 &2.06(0.09) &1.454(255) \\ 
07 &1.78(0.01) &0.1(0.03) &0.478(0.004) &$>10$ &$12^{+15}_{-5}$  &1009(61)  &2775.9 &2.2(0.1) &1.300(252) \\ 
08 &1.78(0.01) &0.13(0.03) &0.485(0.004) &7(5) &$7^{+5}_{-2}$  &968(35)  &2766.7 &2.08(0.07) &1.399(251) \\ 
09 &1.87(0.01) &0.1(0.03) &0.455(0.004) &4(3) &$4^{+2}_{-1}$  &800(14)  &2467.9 &1.72(0.03) &1.054(259) \\ 
10 &1.9(0.01) &0.08(0.03) &0.458(0.004) &4(3) &$3.7^{+3}_{-0.9}$  &780(13)  &2454.4 &1.68(0.03) &1.033(242) \\ 
11 &1.92(0.01) &0.09(0.03) &0.402(0.003) &3(2) &$2.9^{+1}_{-0.5}$  &670(7)  &2128.7 &1.44(0.01) &1.206(256) \\ 
12 &1.93(0.01) &0.06(0.03) &0.376(0.003) &4(4) &$4^{+5}_{-1}$  &630(11)  &2009.5 &1.36(0.02) &1.137(249) \\ 
13 &1.92(0.01) &0.07(0.03) &0.359(0.003) &4(4) &$4^{+6}_{-1}$  &609(13)  &1926.6 &1.31(0.03) &1.065(247) \\ 
\hline
    \noalign{\smallskip}
    \end{tabular}
    }
    \end{flushleft}
(*) gti with biased exposure, the affected data are not taken into account in the analyses.\\
The second third and forth columns report the best fit estimate for the model
in Eq. \ref{eq-lp}. The fifth column reports the value of the SED peak analytically  estimated
from Eq. \ref{eq-lp} according to the best fit results. The sixth and seventh columns report the
$E_p$ and $S_p$ best fit estimates using as best fit model Eq. \ref{eq-lpep}. In the eighth column
we report the flux in the 0.3-10.0 keV band, evaluated by Xspec integrating the Eq. \ref{eq-lp} model.
The ninth colum reports the SED peak flux luminosity evaluated as 
$L_p~\simeq~S_p~4\pi~ D_L^2$, where $D_L\simeq 134.1$ Mpc is the luminosity distance.
In the last column we report the reduced $\chi^2$ and the degrees of freedom concerning the  Eq. \ref{eq-lp}  
fit.
    \end{table*}

\setcounter{table}{1} 
     \begin{table*} 
     \caption{\swx~ Spectral analysis of \mrk. \it{continued}} 
     \label{tab-fit} 
     \begin{flushleft} 
     {\small
    \begin{tabular}{llllllllll}
    \hline 
    \hline 
    Orbit  &$a$ &$b$ &K &$E_p*$ &$E_p$ &$S_p$                            &flux 0.3-10 keV                 &$L_p$                     &$\chi^2_r$/dof \\ 
           &          &         &     &keV    &keV   &\tiny{$10^{-12} erg~ cm^{-2} ~s^{-1}$}  &\tiny{$10^{-12}erg~ cm^{-2} ~s^{-1}$} &\tiny{$10^{45}~erg/s$}    &\\ 
    \hline 
    \noalign{\smallskip}
14 &1.96(0.01) &0.06(0.03) &0.329(0.003) &2(1) &$2.1^{+1}_{-0.4}$  &534(6)  &1722.2 &1.15(0.01) &1.277(236) \\ 
15 &1.98(0.01) &0.1(0.03) &0.309(0.003) &1.3(0.3) &$1.3^{+0.2}_{-0.2}$  &496(6)  &1567.0 &1.07(0.01) &1.135(212) \\ 
16 &1.95(0.02) &0.08(0.04) &0.294(0.004) &2(1) &$2.1^{+2}_{-0.4}$  &479(7)  &1531.0 &1.03(0.01) &1.305(162) \\ 
\hline
\noalign {\smallskip \textit{\textbf{ObsId 00030352016 ~~ Date 6/27/2006 ~~ MJD 53913.139963}}} \\ 
01 &1.99(0.01) &0(0.02) &0.253(0.002) &-~- &-~-(-~-)  &-~-(-~-)  &1350.6 &-~-(-~-) &1.001(261) \\ 
02 &1.96(0.01) &-0(0.03) &0.243(0.002) &-~- &-~-(-~-)  &-~-(-~-)  &1324.9 &-~-(-~-) &1.079(247) \\ 
\hline
\noalign {\smallskip \textit{\textbf{ObsId 00219237000 ~~ Date 7/15/2006 ~~ MJD 53931.277331}}} \\ 
01 &1.65(0.02) &0.14(0.04) &0.509(0.007) &$>10$ &-~-(-~-)  &-~-(-~-)  &3193.8 &-~-(-~-) &1.048(171) \\ 
02 &1.63(0.02) &0.17(0.04) &0.543(0.006) &13(11) &$13^{+10}_{-4}$  &1393(105)  &3406.2 &3(0.2) &1.028(208) \\ 
03 &1.66(0.02) &0.17(0.04) &0.534(0.006) &9(8) &$9^{+7}_{-3}$  &1249(80)  &3240.9 &2.7(0.2) &0.979(191) \\ 
04 &1.72(0.02) &0.13(0.04) &0.507(0.006) &$>10$ &$12^{+16}_{-5}$  &1142(90)  &3016.7 &2.5(0.2) &1.009(191) \\ 
05 &1.7(0.02) &0.18(0.04) &0.498(0.006) &6(4) &$6^{+4}_{-2}$  &1050(44)  &2906.0 &2.26(0.09) &1.017(174) \\ 
06 &1.66(0.02) &0.13(0.03) &0.48(0.005) &$>10$ &$19^{+28}_{-8}$  &1275(134)  &3008.1 &2.7(0.3) &0.960(228) \\ 
\hline
    \noalign{\smallskip}
    \end{tabular}
	}
    \end{flushleft}
(*) gti with biased exposure, the affected data are not taken into account in the analyses.\\
The second third and forth columns report the best fit estimate for the model
in Eq. \ref{eq-lp}. The fifth column reports the value of the SED peak analytically  estimated
from Eq. \ref{eq-lp} according to the best fit results. The sixth and seventh columns report the
$E_p$ and $S_p$ best fit estimates using as best fit model Eq. \ref{eq-lpep}. In the eighth column
we report the flux in the 0.3-10.0 keV band, evaluated by Xspec integrating the Eq. \ref{eq-lp} model.
The ninth colum reports the SED peak flux luminosity evaluated as 
$L_p~\simeq~S_p~4\pi~ D_L^2$, where $D_L\simeq 134.1$ Mpc is the luminosity distance.
In the last column we report the reduced $\chi^2$ and the degrees of freedom concerning the  Eq. \ref{eq-lp}  
fit.
    \end{table*}

\setcounter{table}{2} 
     \begin{table*}
    \caption{\swx~ Orbit merged spectral analysis of \mrk.}
     \label{tab-fit-1} 
     \begin{flushleft} 
     {\small
    \begin{tabular}{llllllllll}
    \hline 
     \hline 
     Interval &$a$ &$b$ &$K$ &$E_p*$ &$E_p$ &$S_p$                            &flux 0.3-10 keV                &$L_p$                       &$\chi^2_r$/dof \\ 
           &          &         &     &keV    &keV   &\tiny{$10^{-12} erg~ cm^{-2} ~s^{-1}$}  &\tiny{$10^{-12}erg~ cm^{-2} ~s^{-1}$} &\tiny{$10^{45}~erg/s $}     &\\ 
    \hline 
    \noalign{\smallskip}
\noalign {\smallskip \textit{\textbf{ObsId 00206476000 ~~ Date 4/22/2006 ~~ MJD 53847.252792}}} \\ 
01 &1.68(0.009) &0.11(0.02) &0.349(0.002) &27(21) &$26^{+19}_{-8}$  &941(60)  &2181.3 &2(0.1)
&1.275(398) \\ 
\textbf{01 X+B} &1.68(0.01) &0.12(0.02)  &0.351(0.002) & &20($^{+10}_{-6}$) & & &  &1.299(401) \\
\hline
02 &1.76(0.01) &0.18(0.02) &0.289(0.002) &5(1) &$4.7^{+1}_{-0.6}$  &558(9)  &1620.6 &1.2(0.02)
&1.187(300) \\ 
03 &1.8(0.01) &0.09(0.02) &0.26(0.002) &14(14) &$14^{+15}_{-5}$  &540(27)  &1499.0 &1.16(0.06) &1.079(331) \\ 
04 &1.79(0.01) &0.14(0.03) &0.257(0.002) &5(3) &$5^{+2}_{-1}$  &488(11)  &1434.8 &1.05(0.02) &1.122(254) \\ 
05 &1.804(0.009) &0.13(0.02) &0.281(0.002) &6(2) &$5.7^{+2}_{-1}$  &532(9)  &1570.4 &1.14(0.02) &1.062(344) \\ 
\hline
\noalign {\smallskip \textit{\textbf{ObsId 00030352005 ~~ Date 4/25/2006 ~~ MJD 53850.267940}}} \\ 
01 &1.84(0.01) &0.15(0.02) &0.273(0.002) &3(1) &$3.3^{+0.7}_{-0.4}$  &480(5)  &1466.7 &1.03(0.01) &1.350(288) \\ 
02 &1.859(0.008) &0.16(0.02) &0.266(0.001) &2.8(0.5) &$2.8^{+0.3}_{-0.2}$  &459(3)  &1411.6 &0.988(0.006) &1.363(379) \\ 
\hline
\noalign {\smallskip \textit{\textbf{ObsId 00030352006 ~~ Date 4/26/2006 ~~ MJD 53851.204919}}} \\ 
01 &1.824(0.009) &0.16(0.02) &0.289(0.002) &3.6(0.9) &$3.6^{+0.6}_{-0.4}$  &519(5)  &1570.7 &1.12(0.01) &1.416(336) \\ 
\hline
\noalign {\smallskip \textit{\textbf{ObsId 00030352007 ~~ Date 4/26/2006 ~~ MJD 53851.951963}}} \\ 
01 &1.87(0.01) &0.16(0.03) &0.265(0.002) &2.6(0.6) &$2.6^{+0.4}_{-0.3}$  &451(5)  &1387.8 &0.97(0.01) &1.085(258) \\ 
\hline
\noalign {\smallskip \textit{\textbf{ObsId 00030352008 ~~ Date 6/14/2006 ~~ MJD 53900.016100}}} \\ 
01(c) &2.05(0.01) &0.15(0.03) &0.233(0.002) &0.67(0.1) &$0.7^{+0.09}_{-0.1}$  &377(3)  &1102.4
&0.811(0.006) &1.108(333) \\ 
\hline
\noalign {\smallskip \textit{\textbf{ObsId 00030352009 ~~ Date 6/15/2006 ~~ MJD 53901.490493}}} \\ 
01 &1.86(0.01) &0.03(0.02) &0.17(0.001) &$>100$ &-~-(-~-)  &-~-(-~-)  &970.3 &-~-(-~-) &1.191(400) \\ 
02 &1.87(0.01) &0.15(0.03) &0.191(0.001) &2.7(0.8) &$2.7^{+0.5}_{-0.3}$  &326(3)  &1007.0 &0.701(0.006) &1.149(319) \\ 
\hline
\noalign {\smallskip \textit{\textbf{ObsId 00030352010 ~~ Date 6/16/2006 ~~ MJD 53902.025296}}} \\ 
01 &1.858(0.007) &0.1(0.01) &0.226(0.001) &6(2) &$6^{+2}_{-1}$  &409(6)  &1244.5 &0.88(0.01) &1.297(413) \\ 
02 &1.77(0.008) &0.07(0.01) &0.27(0.001) &$>10$ &-~-(-~-)  &-~-(-~-)  &1614.6 &-~-(-~-) &1.227(420) \\ 
03 &1.713(0.008) &0.04(0.02) &0.3(0.001) &$>100$ &-~-(-~-)  &-~-(-~-)  &1927.8 &-~-(-~-) &1.268(423) \\ 
04 &1.766(0.006) &0.04(0.01) &0.274(0.001) &$>100$ &-~-(-~-)  &-~-(-~-)  &1680.5 &-~-(-~-) &1.273(477) \\ 
05 &1.823(0.007) &-0.02(0.01) &0.243(0.001) &-~- &-~-(-~-)  &-~-(-~-)  &1478.1 &-~-(-~-) &1.310(439) \\ 
06 &1.748(0.008) &0.03(0.02) &0.294(0.001) &$>100$ &-~-(-~-)  &-~-(-~-)  &1840.3 &-~-(-~-) &1.320(414) \\ 
\hline
\noalign {\smallskip \textit{\textbf{ObsId 00030352011 ~~ Date 6/18/2006 ~~ MJD 53904.038766}}} \\ 
01 &1.808(0.009) &0.12(0.02) &0.319(0.002) &7(3) &$7^{+3}_{-1}$  &613(13)  &1794.7 &1.32(0.03) &1.355(378) \\ 
02 &1.82(0.01) &0.19(0.02) &0.341(0.002) &3(0.7) &$3^{+0.4}_{-0.3}$  &600(6)  &1814.2 &1.29(0.01) &1.348(315) \\ 
03 &1.932(0.008) &0.12(0.02) &0.293(0.002) &1.9(0.3) &$1.9^{+0.2}_{-0.1}$  &479(3)  &1508.2 &1.031(0.006) &1.331(385) \\ 
04(c) &1.985(0.009) &0.09(0.02) &0.256(0.002) &1.2(0.2) &$1.2^{+0.1}_{-0.1}$  &410(3)  &1299.8
&0.882(0.006) &1.115(325) \\ 
05 &1.893(0.008) &0.07(0.02) &0.277(0.001) &6(4) &$6^{+3}_{-1}$  &485(8)  &1506.5 &1.04(0.02) &1.209(395) \\ 
06 &1.884(0.006) &0.08(0.01) &0.271(0.001) &5(2) &$5^{+2}_{-0.9}$  &475(5)  &1475.1 &1.02(0.01) &1.253(473) \\ 
07 &1.866(0.008) &0.05(0.02) &0.222(0.001) &$>10$ &-~-(-~-)  &-~-(-~-)  &1248.9 &-~-(-~-) &1.264(419) \\ 
08(c) &1.989(0.008) &0.09(0.02) &0.213(0.001) &1.2(0.1) &$1.2^{+0.1}_{-0.1}$  &341(2)  &1082.0
&0.734(0.004) &1.148(392) \\ 
09 &1.875(0.007) &0.01(0.01) &0.273(0.001) &$>100$ &-~-(-~-)  &-~-(-~-)  &1569.5 &-~-(-~-) &1.349(466) \\ 
10 &1.821(0.007) &0.06(0.01) &0.298(0.001) &$>10$ &-~-(-~-)  &-~-(-~-)  &1720.5 &-~-(-~-) &1.295(467) \\ 
\hline
\noalign {\smallskip \textit{\textbf{ObsId 00030352012 ~~ Date 6/20/2006 ~~ MJD 53907.001774}}} \\ 
01 &1.842(0.007) &0.16(0.01) &0.289(0.001) &3.1(0.5) &$3.1^{+0.3}_{-0.2}$  &506(3)  &1545.3 &1.089(0.006) &1.179(419) \\ 
02 &1.871(0.008) &0.11(0.02) &0.228(0.001) &4(1) &$3.8^{+0.9}_{-0.5}$  &398(4)  &1230.8 &0.856(0.009) &1.238(356) \\ 
03 &1.861(0.008) &0.17(0.02) &0.287(0.001) &2.6(0.4) &$2.6^{+0.2}_{-0.2}$  &491(3)  &1509.5 &1.056(0.006) &1.322(382) \\ 
04 &1.857(0.009) &0.17(0.02) &0.331(0.002) &2.6(0.4) &$2.6^{+0.2}_{-0.2}$  &568(4)  &1741.4 &1.222(0.009) &1.253(344) \\ 
05 &1.915(0.009) &0.15(0.02) &0.318(0.002) &1.9(0.3) &$1.9^{+0.1}_{-0.1}$  &522(4)  &1623.3 &1.123(0.009) &1.283(328) \\ 
\hline
    \noalign{\smallskip}
    \end{tabular}
    }
    \end{flushleft}
(c) Orbit with a strong cooling contamination.\\
The second third and forth column report the best fit estimate for the model
in Eq. \ref{eq-lp}. The fifth column reports the value of the SED peak analytically  estimated
from Eq. \ref{eq-lp} according to the best fit results. The sixth and seventh columns report the
$E_p$ and $S_p$ best fit estimates using as best fit model Eq. \ref{eq-lpep}. In the eighth column
we report the flux in the 0.3-10.0 keV band, evaluated by X-spec integrating the Eq. \ref{eq-lp} model.
The ninth colum reports the SED peak flux luminoisty evaluated as 
$L_p~\simeq~S_p~4\pi~ D_L^2$, where $D_L\simeq 134.1$ Mpc is the luminosity distance.
In the last column we report the reduced $\chi^2$ and the degrees of freedom concerning the  Eq. \ref{eq-lp}  
fit.
    \end{table*}

\setcounter{table}{1} 
     \begin{table*} 
     \caption{\swx~ Orbit merged spectral analysis of \mrk. \it{continued}} 
     \label{tab-fit} 
     \begin{flushleft} 
     {\small
    \begin{tabular}{llllllllll}
    \hline 
    \hline 
     Interval &$a$ &$b$ &$K$ &$E_p*$ &$E_p$ &$S_p$                            &flux 0.3-10 keV                 &$L_p$                     &$\chi^2_r$/dof \\ 
           &          &         &     &keV    &keV   &\tiny{$10^{-12} erg~ cm^{-2} ~s^{-1}$}  &\tiny{$10^{-12}erg~ cm^{-2} ~s^{-1}$} &\tiny{$10^{45}~erg/s$}    &\\ 
    \hline 
    \noalign{\smallskip}
\noalign {\smallskip \textit{\textbf{ObsId 00030352013 ~~ Date 6/22/2006 ~~ MJD 53908.050644}}} \\ 
01 &1.942(0.007) &0.17(0.01) &0.3(0.001) &1.5(0.1) &$1.47^{+0.06}_{-0.05}$  &485(3)  &1493.6 &1.044(0.006) &1.261(393) \\ 
02(c) &1.982(0.008) &0.12(0.02) &0.255(0.001) &1.18(0.1) &$1.18^{+0.07}_{-0.07}$  &408(2)  &1272.0
&0.878(0.004) &1.247(354) \\ 
03 &1.93(0.01) &0.04(0.03) &0.243(0.002) &$>10$ &-~-(-~-)  &-~-(-~-)  &1323.0 &-~-(-~-) &1.119(244) \\ 
04 &1.81(0.01) &0.03(0.02) &0.281(0.002) &$>100$ &-~-(-~-)  &-~-(-~-)  &1665.7 &-~-(-~-) &1.403(278) \\ 
05 &1.805(0.009) &0.09(0.02) &0.251(0.001) &14(12) &$14^{+13}_{-5}$  &520(23)  &1446.2 &1.12(0.05) &1.268(351) \\ 
06 &1.87(0.006) &0.02(0.01) &0.242(0.001) &$>100$ &-~-(-~-)  &-~-(-~-)  &1380.5 &-~-(-~-) &1.263(455) \\ 
\hline
\noalign {\smallskip \textit{\textbf{ObsId 00030352014 ~~ Date 6/23/2006 ~~ MJD 53909.394900}}} \\ 
01 &1.74(0.01) &0.02(0.02) &0.326(0.003) &$>100$ &-~-(-~-)  &-~-(-~-)  &2060.1 &-~-(-~-) &1.163(281) \\ 
02 &1.67(0.01) &0.06(0.03) &0.372(0.003) &$>100$ &-~-(-~-)  &-~-(-~-)  &2435.1 &-~-(-~-) &1.240(255) \\ 
03 &1.66(0.01) &0.02(0.03) &0.411(0.004) &$>100$ &-~-(-~-)  &-~-(-~-)  &2783.6 &-~-(-~-) &1.110(276) \\ 
04 &1.69(0.01) &0.04(0.03) &0.398(0.003) &$>100$ &-~-(-~-)  &-~-(-~-)  &2603.6 &-~-(-~-) &1.234(265) \\ 
05 &1.63(0.01) &0.08(0.03) &0.436(0.004) &$>100$ &-~-(-~-)  &-~-(-~-)  &2938.2 &-~-(-~-) &1.357(284)
\\ 
\textbf{05 X+B} &1.61(0.01) &0.13(0.02)  &0.434(0.004) &   &34($^{+22}_{-11}$) &  &  &   
&1.399(287)\\
\hline
06 &1.67(0.01) &0.06(0.02) &0.434(0.003) &$>100$ &-~-(-~-)  &-~-(-~-)  &2845.1 &-~-(-~-) &1.079(291) \\ 
07 &1.68(0.01) &0.07(0.03) &0.444(0.004) &$>100$ &-~-(-~-)  &-~-(-~-)  &2870.2 &-~-(-~-) &1.045(282) \\ 
08 &1.76(0.01) &0.07(0.03) &0.393(0.003) &$>10$ &-~-(-~-)  &-~-(-~-)  &2366.8 &-~-(-~-) &1.098(255) \\ 
09 &1.8(0.01) &0.03(0.02) &0.383(0.003) &$>100$ &-~-(-~-)  &-~-(-~-)  &2302.1 &-~-(-~-) &1.041(286) \\ 
10 &1.79(0.01) &0.06(0.03) &0.384(0.003) &$>10$ &-~-(-~-)  &-~-(-~-)  &2259.2 &-~-(-~-) &1.025(249) \\ 
\hline
\noalign {\smallskip \textit{\textbf{ObsId 00215769000 ~~ Date 6/23/2006 ~~ MJD 53909.671942}}} \\ 
01 &1.64(0.01) &0.05(0.02) &0.386(0.003) &$>100$ &-~-(-~-)  &-~-(-~-)  &2617.0 &-~-(-~-) &1.099(310) \\ 
\hline
\noalign {\smallskip \textit{\textbf{ObsId 00030352015 ~~ Date 6/24/2006 ~~ MJD 53910.068887}}} \\ 
01 &1.844(0.007) &0.04(0.01) &0.422(0.002) &$>100$ &-~-(-~-)  &-~-(-~-)  &2434.7 &-~-(-~-) &1.317(418) \\ 
02 &1.778(0.008) &0.12(0.02) &0.479(0.002) &8(4) &$8^{+3}_{-2}$  &970(22)  &2752.4 &2.09(0.05) &1.474(399) \\ 
03 &1.832(0.009) &0.09(0.02) &0.43(0.002) &9(7) &$9^{+6}_{-3}$  &829(24)  &2422.2 &1.78(0.05) &1.181(354) \\ 
04 &1.907(0.005) &0.08(0.01) &0.398(0.001) &4(1) &$3.6^{+0.7}_{-0.4}$  &675(4)  &2131.8 &1.452(0.009) &1.259(482) \\ 
05(c) &1.964(0.009) &0.08(0.02) &0.313(0.002) &1.7(0.4) &$1.7^{+0.2}_{-0.2}$  &505(3)  &1618.7
&1.087(0.006) &1.367(322) \\ 
\hline
\noalign {\smallskip \textit{\textbf{ObsId 00030352016 ~~ Date 6/27/2006 ~~ MJD 53913.139963}}} \\ 
01 &1.973(0.008) &0(0.02) &0.248(0.001) &-~- &-~-(-~-)  &-~-(-~-)  &1338.5 &-~-(-~-) &1.072(339) \\ 
\hline
\noalign {\smallskip \textit{\textbf{ObsId 00219237000 ~~ Date 7/15/2006 ~~ MJD 53931.277331}}} \\ 
01 &1.65(0.01) &0.17(0.02) &0.532(0.004) &11(5) &$11^{+4}_{-2}$  &1299(50)  &3288.2 &2.8(0.1)
&1.147(322) \\ 
 \textbf{01 X+B} &1.64(0.01) &0.20(0.02)  &0.535(0.05) & &8($^{+2}_{-1}$) & & &  &1.235(325) \\
\hline
02 &1.684(0.009) &0.16(0.02) &0.504(0.003) &9(4) &$9^{+3}_{-2}$  &1150(32)  &3035.6 &2.47(0.07) &1.136(369) \\ 
\hline
    \noalign{\smallskip}
    \end{tabular}
    }
    \end{flushleft}
(c) Orbit with a strong cooling contamination.\\
The second third and forth column report the best fit estimate for the model
in Eq. \ref{eq-lp}. The fifth column reports the value of the SED peak analytically  estimated
from Eq. \ref{eq-lp} according to the best fit results. The sixth and seventh columns report the
$E_p$ and $S_p$ best fit estimates using as best fit model Eq. \ref{eq-lpep}. In the eighth column
we report the flux in the 0.3-10.0 keV band, evaluated by X-spec integrating the Eq. \ref{eq-lp} model.
The ninth colum reports the SED peak flux luminoisty evaluated as 
$L_p~\simeq~S_p~4\pi~ D_L^2$, where $D_L\simeq 134.1$ Mpc is the luminosity distance.
In the last column we report the reduced $\chi^2$ and the degrees of freedom concerning the  Eq. \ref{eq-lp}  
fit.
    \end{table*}

\bibliographystyle{aa} 
\bibliography{bibtex.bib} 
\end{document}